\documentclass[prd,aps,twocolumn,a4paper,showkeys,nofootinbib,notitlepage]{revtex4-1}

\usepackage{graphicx,psfrag}
\usepackage{mathrsfs}
\usepackage{amsmath,amsfonts,amssymb}
\usepackage{multirow}
\usepackage{comment}
\usepackage{hyperref}
\usepackage{enumitem}
\usepackage{tcolorbox}
\usepackage{xspace}
\usepackage{enumerate}
\usepackage{subcaption}
\usepackage{xcolor}
\usepackage{algorithm}
\usepackage{algpseudocode}

\newcommand{\be}{\begin{equation}}
\newcommand{\ee}{\end{equation}}
\newcommand{\bea}{\begin{eqnarray}}
\newcommand{\eea}{\end{eqnarray}}
\newcommand{\bel}{\begin{align}}
\newcommand{\eel}{\end{align}}

\newcommand{\tGRAthena}{\texttt{GR-Athena++}}
\newcommand{\GRAthena}{\tGRAthena\xspace}

\newcommand{\tAthena}{\texttt{Athena++}}
\newcommand{\Athena}{\tAthena\xspace}

\newcommand{\tEigen}{\texttt{Eigen}}
\newcommand{\Eigen}{\tEigen\xspace}

\def\p{\partial}

\def\GMc2{{\rm G M_{\odot} c^{-2}}}

\def\eps{\epsilon}
\def\eps{\epsilon}

\def\kt2{\kappa^\text{T}_2}

\def\kt2{\kappa^\text{T}_2}

\def\2nd{2^\mathrm{nd}}
\def\4th{4^\mathrm{th}}
\def\6th{6^\mathrm{th}}
\def\8th{8^\mathrm{th}}

\def\eg{\textit{e.g.}}

\newcommand{\obar}[1]{\overline{#1}}

\def\z4c{$\mathrm{Z}4\mathrm{c}$}
\def\z4oc{$\mathrm{Z}4(\mathrm{c})$}
\def\z4{$\mathrm{Z}4$}
\def\ccz4{$\mathrm{CCZ}4$}

\newcommand{\MeshBlock}{\texttt{MeshBlock}}

\newcommand{\MeshBlocks}{\texttt{MeshBlock}s}
\def\p{\partial}

\def\GMc2{{\rm G M_{\odot} c^{-2}}}

\def\eps{\epsilon}

\def\ft{{\texttt{f2~}}}
\def\ff{{\texttt{f4~}}}

\usepackage{pifont} 

\usepackage{color}
\definecolor{cyan}{rgb}{0,0.9,0.9}
\definecolor{orange}{rgb}{0.9,0.5,0}
\definecolor{magenta}{rgb}{1,0,1}
\definecolor{purple}{rgb}{0.8,0.4,0.8}
\definecolor{gray}{rgb}{0.8242,0.8242,0.8242}
\definecolor{light-gray}{gray}{0.95}

\begin{document}

\title{Subgrid Modelling for Relativistic Magnetohydrodynamics with Machine Learning }

\author{William \surname{Cook}}
\affiliation{Theoretisch-Physikalisches Institut, Friedrich-Schiller-Universit{\"a}t Jena, 07743, Jena, Germany}

\author{Sebastiano \surname{Bernuzzi}}
\affiliation{Theoretisch-Physikalisches Institut, Friedrich-Schiller-Universit{\"a}t Jena, 07743, Jena, Germany}

\date{\today}

\begin{abstract}
  Resolving the impact of magnetic field instabilities in triggering small scale turbulent flow and the associated rearrangement of the field is of critical importance in understanding multimessenger observables in binary neutron star mergers, and angular momentum transport in neutron stars and accretion disks. Direct simulation of these instabilities are unfeasible, however large-eddy simulations can incorporate the impact of this turbulence with a subgrid model. 
   We present the first machine-learning-based subgrid model for special relativistic magnetohydrodynamics, trained using a neural network. We demonstrate its performance in online simulations of the 3D Kelvin-Helmholtz instability through both a priori and a posteriori tests.
  Evaluated in a low resolution simulation, our model captures magnetic field amplification of a simulation at 4 times the resolution with a speed-up of a factor 44. This demonstrates the applicability of such methods in general relativistic simulations of  neutron star mergers and other scenarios.
\end{abstract}


\maketitle

\section{Introduction}
\label{sec:intro}

Astrophysical fluids typically have incredibly high Reynolds numbers of order $\mathrm {Re}\gtrsim 10^{10}$,
and are thus susceptible to the onset of turbulent flow \cite{Radice:2024gic}. In the context of Binary Neutron Star (BNS)
 Mergers, turbulent flow is triggered by the onset of fluid and magnetic field instabilities such as the
Kelvin-Helmholtz Instability (KHI) \cite{Price:2006fi} and the Magneto-Rotational Instability (MRI) \cite{Balbus:1991a} triggered as the stars first merge and in the post merger remnant or accretion disk respectively. The latter instability 
also plays a key role in angular momentum transport in an accretion disk around a black hole. 
The precise structure of the magnetic field in the post merger remnant is unknown, but is driven by this turbulent flow, and is believed to rearrange into a large scale structure that emerges form the remnant, capable of powering the launching of an astrophysical jet \cite{Piran:2004ba}. This jet is believed to generate gamma ray bursts \cite{Blinnikov:1984a,Paczynski:1986px,Goodman:1986a,Eichler:1989ve,Narayan:1992iy}, such as that observed coincident with the BNS merger observed in the gravitational wave spectrum, GW170817 \cite{LIGOScientific:2017vwq,LIGOScientific:2017ync,LIGOScientific:2017zic,Goldstein:2017mmi,Savchenko:2017ffs}.

Including the impact of the magnetic field in numerical simulations is vital to accurately predict multi-messenger observables from BNS mergers \cite{Palenzuela:2021gdo,Hayashi:2024jwt,Bamber:2024wqr,Musolino:2024sju,Most:2025kqf,Gutierrez:2025gkx,Cook:2025frw,Rainho:2025ykl,Kiuchi:2026pgb,Wen:2026mpb} and in isolated NS spacetimes \cite{Tsokaros:2021pkh,Sur:2021awe,Cook:2025zzy,Pinas:2025bpq,Capobianco:2026ots}.  In a numerical simulation of a BNS, fully modelling this turbulent physics, and capturing the nature of the magnetic field, requires resolving a turbulent cascade that may extend down to $\mathcal{O}(1\mathrm{cm})$ \cite{Radice:2024gic}. Such length scales are impractical to resolve in direct numerical simulations (DNS) that also hope to capture ejecta and gravitational waves on length scales of $\mathcal{O}(1000\mathrm{km})$, and also widely span the possible parameter space. Recent work has shown however that such structure may be captured by zoom-in style simulations that stitch together evolutions at multiple length scales \cite{Gutierrez:2026ngt}, or by short lived incredibly high resolution simulations \cite{Kiuchi:2026pgb}. 

The impact of this small scale physics can be included into these large scale simulations however through the use of Large Eddy Simulations (LES). LES treat the evolution equations of  (magneto-)hydrodynamics ((M)HD) in finite volume form as the equations for ``filtered'' quantities, with the finite discretisation length and volume averaging operation providing this implicit filter. These equations capture the evolution of eddies in the turbulent fluid on length scales larger than the filter length scale; while the impact of smaller scale eddies are captured by non-closed terms in the equations arising from the non-commutation of the filtering operation with non-linearities in the flux functions.

LES then require the construction of a ``subgrid model'' to provide a closure of the equations. Such models have been constructed in a physics motivated manner by Smagorinsky \cite{Smagorinsky:1963a}, developed to include dynamical information \cite{Germano:1990a},  and through an analytic expansion of the equations in the gradient approach \cite{Yeo:1987a,Clark:1979a}. Extensions of such approaches to include magnetic fields have also been developed \cite{Theobald:1994a,Mueller:2002a,Grete:2016a,Grete:2016b,Vlaykov:2016a}.

The application of such methods to astrophysical contexts is summarised in modern reviews such as \cite{Miesch:2015a,Schmidt:2015a}, but here we focus on the relatively few applications to our area of interest, compact object merger and evolution. Such methods have been applied in the context of BNS mergers, initially by \cite{Radice:2017zta}, utilising the Smagorinsky subgrid model \cite{Smagorinsky:1963a}, with subsequent relativistic models based on the gradient approach developed \cite{Vigano:2019a,Carrasco:2019uzl,Vigano:2020ouc}.  A complementary approach is taken by the inclusion of viscous GRHD with a simplified Israel-Stewart method \cite{Shibata:2017jyf}, or through the BDNK approach  \cite{Pandya:2021ief,Shum:2025jnl}. Effective models of magnetic field growth have also been incorporated through dynamo modelling \cite{Most:2023sme}.

Within the fluid dynamics community machine learning (ML) techniques have been used to model turbulence \cite{Ling:2016a} and build such subgrid models in non-relativistic contexts \cite{Beck:2018a,Wang:2018a}. It has been proposed recently by \cite{Rosofsky:2020a} that such techniques may also be used in magnetohydrodynamics (MHD), which may be applied to relativistic physics, and it is this approach that we follow in this work.

ML techniques have already been leveraged in the context of relativistic physics, with a recent comprehensive review of the field given in \cite{Huerta:2026nls}. Applications of these techniques have come in detecting compact binary mergers and performing gravitational wave parameter estimation \cite{George:2017pmj,George:2017vlv,Chua:2018woh} and in building physics informed neural operators to solve the MHD equations \cite{Rosofsky:2022lgb,Rosofsky:2023dtc,Kacmaz:2025djp}. Recent work has also demonstrated the first application of neural networks (NN) in an online simulation of relativistic hydrodynamics,  used to evaluate both the fluid equation of state, and the conserved-to-primitive variable conversion \cite{Dieselhorst:2021zet}. Such methods have subsequently been tested for more complicated equation of state (EOS) \cite{Kacmaz:2024fwa} and in a  BNS simulation \cite{Mudimadugula:2025piz}. Recent work has also explored the use of ML techniques in building subgrid models to capture scale separation in galaxy evolution around super-massive black holes \cite{Bhojwani:2025rtn}.

In this paper we follow the strategy of \cite{Rosofsky:2020a} and perform high resolution simulations of the KHI in both Newtonian and special relativistic (SR) MHD, and train NNs on this data to provide a subgrid model for MHD turbulence. We then demonstrate that these models can be evaluated a posteriori within low resolution simulations of the KHI to match the behaviour of high resolution simulations.

This paper is structured as follows. In Section \ref{sec:meth} we discuss  the 
filtered equations that we solve and the subgrid terms in Newtonian MHD and SRMHD. We then discuss 
the application of ML methods to the problem and discuss the structure of the NNs employed.
In Section \ref{sec:res} we discuss the performance of our subgrid model, both in a priori 
tests of the accuracy of the NN in reproducing test data, and in a posteriori studies, showing the 
ability of the model to accurately model the development of turbulence in online simulations. We show these tests first for Newtonian and then SR simulations.
We summarise our findings in Section \ref{sec:con}.

In the Newtonian case we work in units where $\epsilon_0 = \mu_0 = 1$, and absorb a factor of $4\pi$ into the magnetic field definition, these are the standard units used in Athena++ \cite{Stone:2020}. In special relativity we additionally set $c=1$.

\section{Method}
\label{sec:meth}

\subsection{Filtering}

When evolving equations using a finite volume approach the evolved variables are volume averaged, equivalently ``filtered'', over a given computational cell. The size of this cell provides the smallest length scale that can be directly resolved in the simulation. In a turbulent flow, where eddies may cascade in size to smaller and smaller scales, only the eddies larger than this length scale can be resolved. In the LES approach, rather than viewing the implicit filtering of the fluid as an artefact of the numerical approach, we ask what are the exact equations that the filtered fluid obey. By filtering the evolution equations we obtain a system that describes the  exact evolution of the fluid on length scales longer than the discretisation length of our simulation, before any numerical approximations are made. The filtered fluid variables then directly correspond to the evolved variables in a finite volume simulation. However, the filtration operation does not commute with non-linearities in the fluxes of our evolution equations and the fluxes will depend on the filtration of product terms. These are not evolved variables and require some knowledge of the solution to the equations on the sub-filter scale. The system is therefore not closed, and requires the prescription of these extra ``subgrid terms'' through a ``subgrid model''. Below we derive the filtered evolution equations and associated subgrid terms for both Newtonian and SR MHD, before proceeding to generate such a subgrid model. 

Let us first define the concept of filtering. We follow the standard approach for filtering in turbulent flows, as detailed in \eg \cite{Leonard:1975a,Pope:2000,Sagaut:2006a}.
We define a filtering kernel, $G(r,x)$, where $G$ is normalised such that $\int G(r,x) dr = 1$.

A filtered variable is defined as the convolution with the filtration kernel i.e. 
\begin{eqnarray}
	\obar{f}(x,t) = \int G(r,x)f(x-r,t) dr 	.
\end{eqnarray}
Convolution with the filtration kernel commutes with $t$ and $x$ derivatives, as long as $G$ does not depend on $x$, i.e. it is homogeneous. This is true,  for instance, of a top hat box filter,
\begin{eqnarray}
	G(r) = \frac{1}{\Delta x} H\left(\frac{\Delta x}{2} - |r|\right),
\end{eqnarray}
where $H$ is the Heaviside step function. We can consider this filtering operation to be the same as that which is applied when volume averaging a continuous variable over a computational cell of width $\Delta x$, as in the standard finite volume approach to computational fluid dynamics.

By redefining $r = x - x'$, we recover the slightly more common expression for a filtered variable:
\begin{eqnarray}
	\obar{f}(x,t) &=&\int^{x'_\mathrm{max}}_{x'_\mathrm{min}} G(x-x') f(x',t) dx'
\end{eqnarray}
We define a second filtering operation, the Favre filter. A quantity filtered by the Favre filter is denoted with a tilde, and defined,
\begin{eqnarray}
	\tilde{f} = \frac{\obar{\rho f}}{\obar{\rho}}
\end{eqnarray}
whee $\rho$ is the fluid density. We note therefore trivially that $\obar{\rho f} = \obar{\rho}\tilde{f}$. We will employ this filtering operation in the Newtonian equations following, and define the evolved velocity as the Favre filtered velocity. Consequently, there will be no subgrid terms in the density evolution equation.

\subsection{Newtonian MHD}

In Newtonian MHD the evolution equations for a magnetised fluid read,
\begin{eqnarray}
	\partial_t \rho + \partial_i(\rho v^i) &=&  0 \\
	\partial_t(\rho v^j)  + \partial_i(\rho v^i v^j - B^i B^j \nonumber
	 &&\\ + \delta^{ij} (p + B^2/2)) &=& 0  \\
	\partial_t U  + \partial_i ((U + p + \frac{B^2}{2})v^i - (v_j B^j)B^i)&=&0 \\
	\partial_t B^k + \partial_i(B^kv^i - B^iv^k) &=&0  .
\end{eqnarray}

Recall also the definition of $U$,
\begin{eqnarray}
	U = e + \frac{\rho v^2}{2} + \frac{B^2}{2}
\end{eqnarray}

These are the continuum equations, with no filtering, or discretisation procedures applied. We now filter these equations, integrating against the filtration kernel $G(x-x')$. 
\begin{eqnarray}
	\partial_t \obar{\rho} + \partial_i( \obar{\rho v^i}) &=&  0 \label{eq:NMHD} \\
	\partial_t(\obar{\rho v}^j)  + \partial_i(\obar{\rho v^i v^j} - \obar{B^i B^j} + \delta^{ij} \obar{(p + B^2/2)}) &=& 0  \\
	\partial_t \obar{U}  + \partial_i (\obar{(U + p + \frac{B^2}{2})v^i} - \obar{(v_j B^j)B^i})&=&0 \\
	\partial_t \obar{B^k} + \partial_i(\obar{B^kv^i - B^iv^k}) &=&0  
\end{eqnarray}
Now let us rearrange to form the LES equations, using the Favre filter for the velocity.
\begin{widetext}
\begin{eqnarray}
	\partial_t \obar{\rho} + \partial_i( \obar{\rho} \tilde{v^i}) &=&  0 \\
	\partial_t(\obar{\rho} \tilde{v}^j)  + \partial_i( \obar{\rho} \tilde{v}^i \tilde{v}^j - \obar{B}^i \obar{B}^j + \delta^{ij} (\hat{p} + \obar{B}^2/2) - \tau_\mathrm{S}^{ij}) &=& 0  \\
	\partial_t \obar{U}  + \partial_i ((\obar{U} + \hat{p} + \obar{B}^2)\tilde{v}^i - (\tilde{v}_j \obar{B}^j)\obar{B}^i - \tau_\mathrm{U}^i)&=&0  \\
	\partial_t \obar{B}^k + \partial_i(\obar{B}^k\tilde{v}^i - \obar{B}^i\tilde{v}^k - \tau_\mathrm{B}^{ik}) &=& 0,
\end{eqnarray}
where we define 
\begin{eqnarray}
	\tau_\mathrm{S}^{ij} &=& \left(\obar{\rho}\tilde{v}^i\tilde{v}^j - \obar{\rho v^i v^j}\right) - \left(\obar{B}^i\obar{B}^j - \obar{B^i B^j}\right) + \delta^{ij}\left(\obar{p} - \hat{p} + \frac{\obar{B}^2 - \obar{B^2}}{2}\right) \\
	&=& \obar{\rho}\left(\tilde{v}^i\tilde{v}^j - \widetilde{v^i v^j}\right) - \left(\obar{B}^i\obar{B}^j - \obar{B^i B^j}\right) + \delta^{ij}\left(\obar{p} - \hat{p} + \frac{\obar{B}^2 - \obar{B^2}}{2}\right)\label{eq:sgnewtS} \\
	\tau_\mathrm{U}^i &=& \left(\obar{U}\tilde{v}^i - \obar{Uv^i}\right) + \left(\hat{p}\tilde{v}^i - \obar{pv^i}\right) + \left(\obar{B}^2\tilde{v}^i - \obar{B^2v^i}\right) - 
	\left(\obar{B}^i\tilde{v}_j\obar{B}^j - \obar{B^iv_jB^j}\right)\label{eq:sgnewtU}\\
	\tau_\mathrm{B}^{ik} &=& \left(\obar{B}^k\tilde{v}^i - \obar{B^kv^i}\right) - \left(\obar{B}^i \tilde{v}^k - \obar{B^iv^k}\right),\label{eq:sgnewtB}
\end{eqnarray}
\end{widetext}
where in the second line we have applied the definition of the Favre filter to the product of velocities.

Explicitly, we consider a Gamma law equation of state, with $\Gamma = 4/3$, and so the 
pressure as calculated from the filtered variables is
\begin{eqnarray}
	\hat{p} &=& (\Gamma - 1) \hat{e}\\
	\hat{e} &=& \obar{U} - \frac{\obar{\rho}\tilde{v}^2}{2} - \frac{\obar{B}^2}{2}
\end{eqnarray}

\subsection{Special Relativistic MHD}

For special relativity, conservation of mass, energy momentum and the Maxwell equations in the ideal MHD approximation give the evolution equations for a magnetised fluid,
\begin{eqnarray}
	\p_t D + \p_i\left(Dv^i\right) &=& 0\\
	\p_t S_j + \p_i\left( S_j  v^i  + \delta^i_j\left( p + \frac{b^2}{2}\right)  -\frac{b_jB^i}{W} \right) &=& 0\\
	\p_t\tau + \p_i\left( \left(\tau v^i +(p + \frac{b^2}{2})v^i - \frac{ b^0 B^i}{W}  \right)\right) &=&
       0\\
	\p_tB^k + \p_i\left(  \left( B^k v^i  - B^i v^k\right)\right) &=& 0 
\end{eqnarray}
Here the conserved variables can be written in terms of the primitive variables like so,
\begin{eqnarray}
D &=& \rho W\\
S_j &=& (\rho h + b^2) W^2 v_j - b^0b_j\\
\tau &=& (\rho h + b^2) W^2 - \rho W - (p + b^2/2) - (b^0)^2,
\end{eqnarray}
where the Lorentz factor $W = \frac{1}{\sqrt{1-v^2}}$, and $h = 1+\eps + p/\rho$ is the fluid enthalpy. The magnetic field in the comoving fluid frame, $b$, is related to the magnetic field as seen by the Eulerian observer, $B$, by
\begin{eqnarray}
	b^0 &=& WB^iv_i\\
	b^i &=& \frac{B^i}{W} + WB^jv_j v^i\\
	b^2 &=& \eta_{\mu\nu}b^\mu b^\nu = (B^jv_j)^2 + \frac{B^jB_j}{W^2}
\end{eqnarray}
We filter these equations as above, obtaining
\begin{widetext}
\begin{eqnarray}
	\p_t \obar{D} + \p_i\left(\obar{Dv^i}\right) &=& 0\\
	\p_t \obar{S_j} + \p_i\left( \obar{S_j  v^i}  + \delta^i_j\left( \obar{p} + \obar{\frac{b^2}{2}}\right)  -\obar{\left(\frac{b_jB^i}{W}\right)} \right) &=& 0\\
	\p_t\obar{\tau} + \p_i\left( \left(\obar{\tau v^i} +\left(\obar{p v^i} + \frac{\obar{b^2v^i}}{2}\right) - \obar{\left(\frac{ b^0 B^i}{W}\right)}  \right)\right) &=&
	0\\
	\p_t\obar{B^k} + \p_i\left(  \left( \obar{B^k v^i}  - \obar{B^i v^k}\right)\right) &=& 0 .
\end{eqnarray}
We can then rearrange to form the LES equations,
\begin{eqnarray}
	\p_t \obar{D} + \p_i\left(\obar{D}\hat{v^i} - \tau^i_\mathrm{D}\right) &=& 0\label{eq:srLES_D}\\
	\p_t \obar{S_j} + \p_i\left( \obar{S_j}  \hat{v^i}  + \delta^i_j\left( \hat{p} + \frac{\hat{b^2}}{2}\right)  -\left(\frac{\hat{b_j}\obar{B^i}}{\hat{W}}\right) - \tau^{ij}_\mathrm{S}\right) &=& 0\label{eq:srLES_S}\\
	\p_t\obar{\tau} + \p_i \left(\obar{\tau} \hat{v^i} +\left(\hat{p} \hat{v^i} + \frac{\hat{b^2}\hat{v^i}}{2}\right) - 
	 \frac{ \hat{b^0} \obar{B^i}}{\hat{W}} - 
	\tau_\mathrm{\tau}^i
	 \right) &=&
	0\label{eq:srLES_t}\\
	\p_t\obar{B^k} + \p_i\left(  \left( \obar{B^k} \hat{v^i}  - \obar{B^i} \hat{v^k}\right) - \tau^{ik}_{\mathrm{B}}\right) &=& 0 \label{eq:srLES_B},
\end{eqnarray}
where
\begin{eqnarray}
	\tau^i_\mathrm{D} &=& \obar{D}\hat{v^i} - \obar{Dv^i} \label{eq:sr_sg_1} \\
	\tau^{ij}_\mathrm{S} &=&  \obar{S_j}  \hat{v^i} - \obar{S_j  v^i}  + \hat{p} - \obar{p} + \hat{\frac{b^2}{2}} - \obar{\frac{b^2}{2}} -\left(\left(\frac{\hat{b_j}\obar{B^i}}{\hat{W}}\right)- \obar{\left(\frac{b_jB^i}{W}\right)}   \right)\label{eq:sr_sg_2}\\
	\tau^i_\mathrm{\tau} &=&  \obar{\tau} \hat{v^i} - \obar{\tau v^i}  +  \hat{p} \hat{v^i} -  \obar{p v^i} + \frac{\hat{b^2}\hat{v^i}}{2} -  \frac{\obar{b^2v^i}}{2}  -  \left( \frac{ \hat{b^0} \obar{B^i}}{\hat{W}} -  \obar{\left(\frac{ b^0 B^i}{W}\right)} \right)\label{eq:sr_sg_3}\\
	\tau^{ik}_{\mathrm{B}} &=&  \obar{B^k} \hat{v^i} - \obar{B^k v^i}  - \left( \obar{B^i} \hat{v^k} - \obar{B^i v^k} \right)\label{eq:sr_sg_4}
	\end{eqnarray}
	
\end{widetext}
We note that, by rearranging the flux terms in the energy equation for $\tau$, one can eliminate the subgrid term $\tau_\tau^i$ and write this in terms of the subgrid term appearing in the equation for mass density $\tau_D^i$. We have experimented with this option in our simulations and find no overall difference between these two implementations of the subgrid terms.

\subsection{Model training with Machine Learning}

\subsubsection{Machine Learning Approach} \label{sec:architecture}

The subgrid tensor terms demonstrated in Eqs \ref{eq:sgnewtS}-\ref{eq:sgnewtB} and Eqs \ref{eq:sr_sg_1} - \ref{eq:sr_sg_4} for Newtonian and SR MHD respectively are not closed. 
While filtered individual conserved variables are directly accessible in the simulation, since they are evolved variables, filtered products of variables are not. We therefore provide a closure,
specifying these terms as functions of the fluid state. We take the approach suggested in \cite{Rosofsky:2020a}, and widely applied in the fluid dynamics community, to train a NN on the relationship between the state of the fluid (as inputs to the network) and the subgrid tensors (the outputs). We now outline the explicit procedure we follow, and give concrete examples for the induction equation, and associated subgrid tensor $\tau_B$, specifically for Newtonian physics.

We initially perform a high resolution simulation of turbulent flow, here specifying to a local simulation of the KHI (see initial data below). We denote the conservative and primitive variables obtained in this simulation $\mathbf{q}^{\mathrm{HR}}, \mathbf{w}^{\mathrm{HR}}$. We treat this as the continuum solution, and calculate the product terms that appear in subgrid tensors, e.g. $B^{i,\mathrm{HR}} v^{k,\mathrm{HR}}$. We now filter the high resolution data down to low resolution, $\obar{q} = \obar{\mathbf{q}^{\mathrm{HR}}}$, by averaging neighbouring cells, and filter the product terms also, $\obar{B^iv^k} = \obar{B^{i,\mathrm{HR}} v^{k,\mathrm{HR}}}$. Now we can construct the subgrid tensor at low resolution $\tau_B = \left(\obar{B}^k\tilde{v}^i - \obar{B^kv^i}\right) - \left(\obar{B}^i \tilde{v}^k - \obar{B^iv^k}\right)$. This is then the output on which we train our model, with the inputs given by the filtered individual terms (along with their spatial derivatives). Note the unique feature to Newtonian MHD, that the Favre filter is applied to the individual velocity term $v$. Our NNs are fully connected Multi Layer Perceptrons, which we train using the PyTorch library \citep{Paszke:2019a}.

Once our dataset is constructed we separate off $10\%$ to be used as a validation dataset to track the loss function during training. We use a mean squared error loss function for all models. We perform our training for a maximum of 200 epochs, and implement an early stopping criterion, such that, if for 5 epochs the validation error does not improve by more than $10^{-4}$, the training terminates. The model at the epoch with the lowest validation error is then picked.

\subsubsection{Conserved to Primitive}

In SRMHD the above algorithm must be slightly modified, since the conserved to primitive variable inversion is not linear, and hence does not commute with the filtering operation. The primitive variables that are evaluated in the fluxes of a numerical simulation are calculated from the conserved variables, and are not evolved variables themselves. When we come to discuss filtering therefore, they must be treated differently, since, when fluxes are evaluated within our evolution,  we are using, not the filtered version of the true primitives, but instead, the primitives that are calculated from the filtered conservatives. To make this explicit, let the set of conservatives be denoted $q$, and the primitives, $w$. $w$ can be calculated from $q$ through the non linear relationship $w = g(q)$.
Define $\hat{w} = g(\overline{q})$,
i.e., not the directly filtered primitives (which would be $\overline{w} = \overline{g(q)}$), but the primitives calculated from the filtered conservatives.

The non-trivial conserved-to-primitive makes assembling the training data for the NN slightly  more complicated than for the Newtonian case. Explicitly, returning to the above example of the induction equation, at high resolution the product term is calculated as above, and filtered in the same way. We then filter the conserved variables from our high resolution simulation, and perform a conserved to primitive conversion on those filtered variables, to obtain the ``low resolution'' primitives, rather than directly filtering the ``high resolution'' primitives from the simulation. Hence the subgrid tensor $\tau^{ik}_B = \obar{B^k} \hat{v^i} - \obar{B^k v^i}  - \left( \obar{B^i} \hat{v^k} - \obar{B^i v^k} \right)$.

\subsubsection{Choice of input and outputs for NN} \label{sec:meth_inputs}

Following the approach of \cite{Rosofsky:2020a}, we use as inputs for the NNs the state of the fluid, as well as its spatial derivatives at a given spatial point, with the corresponding output, the subgrid tensor at that point. In \cite{Rosofsky:2020a}, the inputs passed to the NN were not just these variables at a given point, but also those at neighbouring grid points in a stencil of fixed size. In order to restrict the size of the models we train, and keep our online evolution of the model efficient, we do not use a stencil of neighbouring points. In the Newtonian  case we use only the primitive variables as input, since the conserved-to-primitive conversion is trivial, but in the SR case, we supply both the conserved and primitive variables, and their derivatives, as inputs, implicitly encoding the nature of the relationship between the two, and the equation of state, within the network. We train a separate model for each subgrid tensor, with each component of the tensor a different output of the network. During the training process both the inputs and outputs of the training dataset are normalised to a unit normal distribution.

\subsubsection{Optimised evaluation}

The aim of incorporating a subgrid model is to allow computationally inexpensive low resolution simulations to capture physics that would otherwise only be accessible in higher resolution, more expensive simulations. It is therefore necessary that the cost of evaluating the subgrid model is relatively cheap compared to the overall evolution speed, such that the low resolution simulation with the subgrid model activated is still cheaper than the equivalent high resolution simulation with no model. In other applications of NNs within relativistic hydrodynamics evolutions, it has been found that NN may speed up the evolution compared to traditional methods \cite{Dieselhorst:2021zet}, though other applications have found that in full production contexts there is no useful acceleration \cite{Mudimadugula:2025piz}.

Here we attempt to implement our subgrid model in a manner that ensures a fast evaluation during the evolution. Our first goal is to ensure that models stay relatively small. We discuss the full range of hyperparameters we have explored in App. \ref{app:hyp}, and here we note that we target models with 6 or fewer hidden layers ($N_L$), and no more than 64 neurons per layer ($N_n$). This ensures that the matrix multiplications in the model evaluation are relatively cheap. We  directly evaluate the models ``by hand'' during our evolution, loading the model weights and biases, and evaluating the model using the matrix multiplication library \Eigen \cite{eigenweb} to ensure fast, optimised evaluation of the network. We also use the mean and standard deviation of the training data sets to rescale the inputs and outputs of the model.

We evaluate the network at every substage in the multistage time evolution algorithm (RK3), and we enforce that symmetric subgrid terms are equal, i.e. we set $\tau_{ij} = \tau_{ji}$ where appropriate.

\subsection{Initial data \& numerical details}

\subsubsection{Numerical approach}\label{sec:numerics}

Within this paper we consider two test problems, firstly a 2D KHI in Newtonian MHD, and secondly a 3D KHI in SR. For the Newtonian problem we evolve the system using the code \Athena \cite{Stone:2020}, while  the SR case is evolved with \GRAthena \cite{Daszuta:2021ecf,Cook:2023bag,Daszuta:2024chz,Daszuta:2024ucu,Daszuta:2026szb}. The numerical fluxes are constructed using an LLF approximate Riemann solver and PPM reconstruction, and time integration is performed with an RK3 integrator. All numerical grids considered are of uniform resolution, with a box side length of 2 in code units. For the Newtonian case, grid configurations LR and HR have grid spacing $\Delta x = 2 / (256, 512)$ while in the special relativistic case grid spacing $\Delta x = 2 / (128, 256, 512)$ for grid configurations LR, SR, HR respectively (here SR stands for standard resolution).

As can be seen from \eg~Eqs \ref{eq:srLES_D} - \ref{eq:srLES_B}, the subgrid terms appear as additional contributions to the fluxes. We note that there are two options for how to evaluate these terms. Firstly one can follow the approach we take here, evaluating these terms as additional contributions to the physical fluxes, but an alternate approach is to move these terms, with their associated derivatives, to the right hand side and to treat them as source terms.

There are advantages and disadvantages to both approaches. The mass density evolution equation is written as an exact conservation law, which preserves conservation of mass to machine precision, up to the application of a floor, or outflows from a numerical grid. Moving subgrid terms to the RHS breaks this explicit framing of the conservation law, and may lead to violations of mass conservation. \Athena and \GRAthena employ the constrained transport algorithm \cite{Evans:1988a,Gardiner:2007nc} to evolve the induction equation, which relies on writing the flux for the magnetic field as an exact curl, in order to preserve the divergence free condition on the magnetic field to machine precision. Again, introducing source terms on the RHS may introduce monopoles into our evolution, and so we retain the subgrid terms within the flux calculation, rather than introducing them as sources on the RHS.

The cost to be paid for retaining the subgrid terms within the fluxes, comes in constructing the inputs to the NN. The NN depends on both the fluid state and its spatial derivatives, and these must be evaluated at the face centres of the computational grid, as this is where the flux is calculated. The primitive variables are already reconstructed to the face centres as is standard for the construction of the ordinary flux terms, however now we must also reconstruct the derivatives of the primitives to the face centres.

To do this we perform a two step procedure. Firstly, the spatial derivatives of the primitives are calculated at the cell centres using first order finite differencing, with a minmod limiter. Then the cell centred derivatives are reconstructed to the cell interface using donor cell reconstruction. In the implementation of the constrained transport algorithm, the evolution of the induction equation requires the calculation of the electric field at two grid samplings. Firstly, at the face centred sampling, as part of the flux calculation described above, and secondly, at cell centres, from $E = - v \times B$. We also correct this term with the same subgrid term that occurs in the flux of the induction equation, though the input to the NN here is given by the cell centred primitives and their already calculated derivatives, rather than the face centred quantities. Once the subgrid corrected electric field at the faces and cell centres has been calculated, the constrained transport algorithm proceeds as normal, calculating the edge centred electric field and so integrating the magnetic field in time. 

\subsubsection{Newtonian initial data} \label{sec:N_ID}

For initial data for the 2D Newtonian problem we follow the initial data of \cite{Rosofsky:2020a}, reproduced below.
\begin{eqnarray}
	\rho &=& \rho_0 +\rho_1\mathrm{sgn}(y)\tanh\left(\frac{|y| - y_l}{a_l}\right),\\
	v^x &=& v^x_0\mathrm{sgn}(y)\tanh\left(\frac{|y| - y_l}{a_l}\right) + \delta v^x \sin\left(2\pi n_x y\right)\nonumber\\&&\\
	v^y &=&  \mathrm{sgn}(y)\delta v^y \sin\left(2\pi n_y x\right)\exp\left(-\frac{(|y| - y_l)^2}{\sigma_y^2}\right)\nonumber\\&&\\
	B^x &=& B_0\\
	B^y&=&  0\\
	p &=& p_0,
\end{eqnarray}
with chosen parameter values $p_0 = 1, a_l = 0.01, \sigma = 0.1, y_l = 0.25, B_0 = 0.001, \rho_0 = 1.5, \rho_1 = -0.5, v^x_0 = 0.5, \delta v^x = 0.01, n_x = 4, n_y = 7, \delta v^y = 0.2$.

\subsubsection{Special Relativistic initial data} \label{sec:SR_ID}

For a fully turbulent 3D KHI in special relativity we follow the initial data specified by \cite{Vigano:2020ouc}. We initialise the primitive variables as follows:
\begin{eqnarray}
	\rho &=& \rho_0 +\rho_1\mathrm{sgn}(y)\tanh\left(\frac{|y| - y_l}{a_l}\right),\\
	v^x &=& v^x_0\mathrm{sgn}(y)\tanh\left(\frac{|y| - y_l}{a_l}\right) + \delta v^x \sin\left(\frac{2\pi n_x x}{L_x}\right)\nonumber\\&&\\
	v^y &=& v^y_0\mathrm{sgn}(y)\tanh\left(\frac{|y| - y_l}{a_l}\right) \nonumber \\ &&+ \delta v^y \sin\left(\frac{2\pi n_y y}{L_y}\right)\mathrm{sgn}(y)\exp\left(-\frac{(|y| - y_l)^2}{\sigma_y^2}\right)\nonumber\\&&\\
	v^z  &=& v^z_0\mathrm{sgn}(y)\exp\left(-\frac{(|y| - y_l)^2}{\sigma_y^2}\right) + \delta v^z \sin\left(\frac{2\pi n_z z}{L_z}\right)\nonumber \\&&\\
	B^x &=& B_0\\
	B^y&=& B^z = 0\\
	p &=& p_0,
\end{eqnarray}
where the chosen parameter values are $B_0 = 0.001, p_0 = 1, \rho_0 = 2, \rho_1 = 1, y_l = 0.25, a_l = 0.01, v^i_0 = 0.5, \delta v^x = 0.01, \delta v^y = 0.1, \delta v^z =0.01, n_x=11, n_y = 7, n_z = 5, \sigma^2_y = 0.01, \sigma^2_z = 0.1 $

\section{Results}
\label{sec:res}

\subsection{Newtonian Kelvin-Helmholtz Instability in 2D}\label{sec:res:newt}

\subsubsection{Training Dataset construction}

We evolve the initial data detailed in Sec. \ref{sec:N_ID} up to $t=100$, ensuring that, for an extended period in the simulation, the initial configuration has been totally washed out, and that a fully turbulent evolution has developed, at two resolutions, LR and HR. We only use data from after this burn in period in the training dataset. Following the filtering procedure described in Sec. \ref{sec:architecture}, we calculate the subgrid terms on the HR dataset and filter this data down to the LR resolution. We then perform the data augmentation procedure detailed in App. \ref{app:hyp} to construct our training dataset of inputs (detailed in Sec. \ref{sec:meth_inputs}) and the corresponding subgrid terms. 

\subsubsection{Model training and a priori testing}

\begin{figure}[t]
  \centering 
  \includegraphics[width=0.49\textwidth]{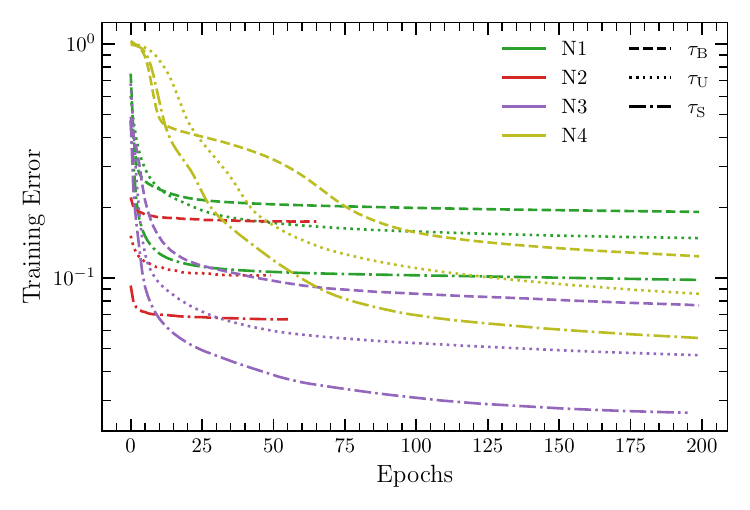}
  \caption{ The error of the NN during training. We show 4 models with different hyperparameters, selected by a posteriori criteria discussed in Sec. \ref{sec:newt:apost}. Colours represent different models, while linestyles represent the three networks trained for each model, corresponding to $\tau_\mathrm{B}, \tau_\mathrm{U}, \tau_\mathrm{S}$ (Eqs \ref{eq:sgnewtS},\ref{eq:sgnewtU},\ref{eq:sgnewtB}).} 
  \label{fig:newt2d_training}
\end{figure}

We train a number of NNs on this training dataset exploring a range of hyperparameters as detailed in App. \ref{app:hyp}. In Fig. \ref{fig:newt2d_training} we demonstrate the mean square error loss of four of our networks during training. We demonstrate model architectures which have been selected with a posteriori criteria defined below, with the model hyperparameters detailed in Tab. \ref{tab:newt2dhyp}. The stopping criteria for our training is detailed in Sec. \ref{sec:architecture}, and we see that, for instance, model N2 terminates training comparatively early, after $\sim 60$ epochs, while all other models continue to train until the maximum 200 epochs. Despite this, the error of this model is lower than that of model N1, while the lowest error is found for model N3 across the 3 subgrid models constructed for the 3 subgrid tensors.

We validate the a priori performance of our models by evaluating the model on a testing dataset drawn from a timeslice of the HR simulation not used to generate the training dataset. The testing dataset consists of pairs of the fluid state at the filtered resolution (the input to the network), and the corresponding subgrid tensor values, calculated at higher resolution, and then filtered (the outputs) . We compare the model's predicted values of the subgrid tensors to the exact values in the dataset. 

\begin{figure*}[t]
  \centering 
  \includegraphics[width=0.32\textwidth]{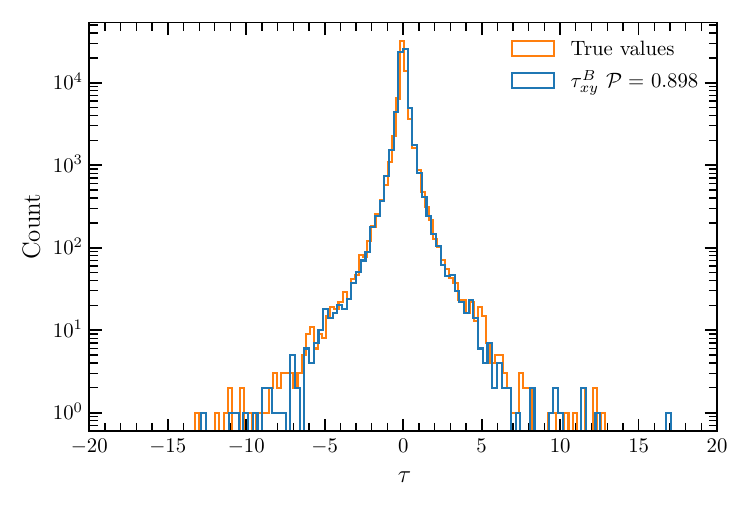}
  \includegraphics[width=0.32\textwidth]{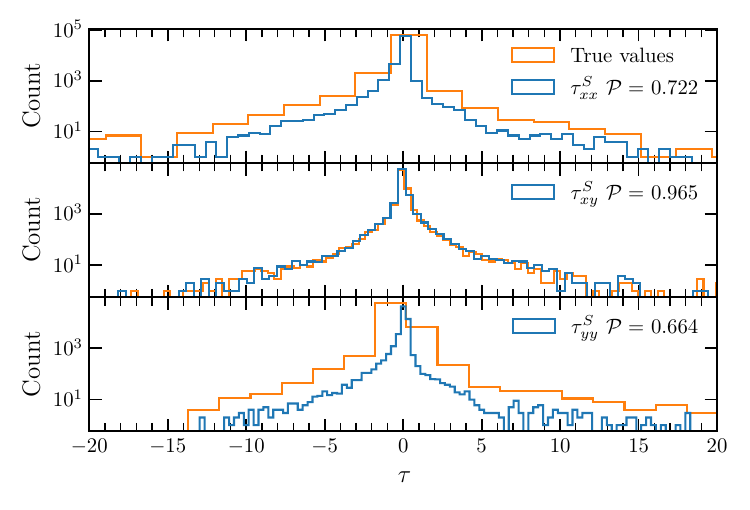}
  \includegraphics[width=0.32\textwidth]{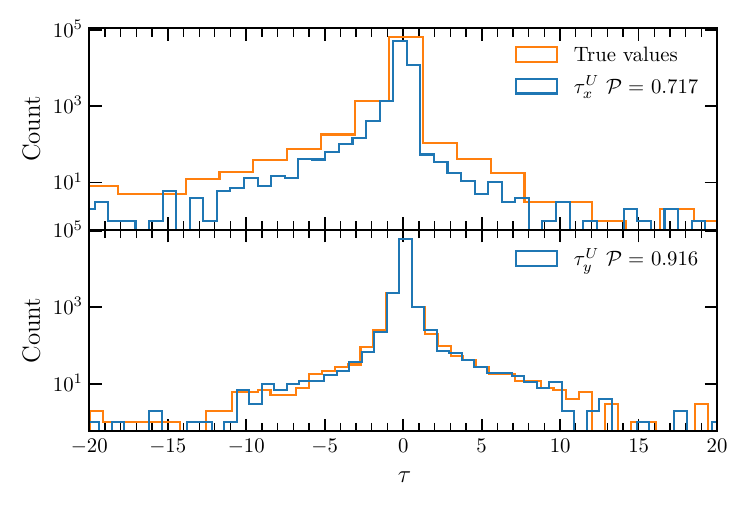}
  \caption{Histogram of the subgrid tensors in the testing dataset (orange) and the predicted values of those tensors by the NN model N1. On the x-axis we show the value of the normalised subgrid tensor. On the y-axis we show the unnormalised count of points in the bins. Left, middle and right panels show the subgrid tensors $\tau_B,\tau_S,\tau_U$ respectively, while subpanels show different components of the tensor, indicated in the legend. In the legend we display the Pearson correlation coefficient between the two datasets. } 
  \label{fig:newt2d_apriori_histo}
\end{figure*}

In Figure \ref{fig:newt2d_apriori_histo} we show the distribution of values of each subgrid tensor in the testing dataset, and the corresponding distribution predicted by network N1. We also calculate the Pearson correlation coefficient between the two datasets, denoted $\mathcal{P}$. A value of 1 corresponds to a perfect correlation, which we would expect for a model that perfectly generated the correct subgrid tensor values, while a value of 0 corresponds to uncorrelated data, which would denote a model generating random values. We see that our model performs well in this a priori test, with the predicted distributions matching the expected values well. We find also that the correlation coefficient is well above zero, and over 0.7 for all but one component. The other models considered, N2, N3, N4 show very similar performance to model N1 in all components. The worst performing component in this test is $\tau_{yy}^S$, which we ascribe to the comparatively small value of the momentum in the $y$ direction, compared to the $x$ direction due to the setup of our test problem. We may be able to further reduce this error in the future by training separate models for each component of the subgrid tensor, though the impact on performance in online evaluation of the network may make this improvement unfeasible.

\subsubsection{A posteriori testing} \label{sec:newt:apost}

\begin{figure*}[t]
  \centering 
  \includegraphics[width=0.99\textwidth]{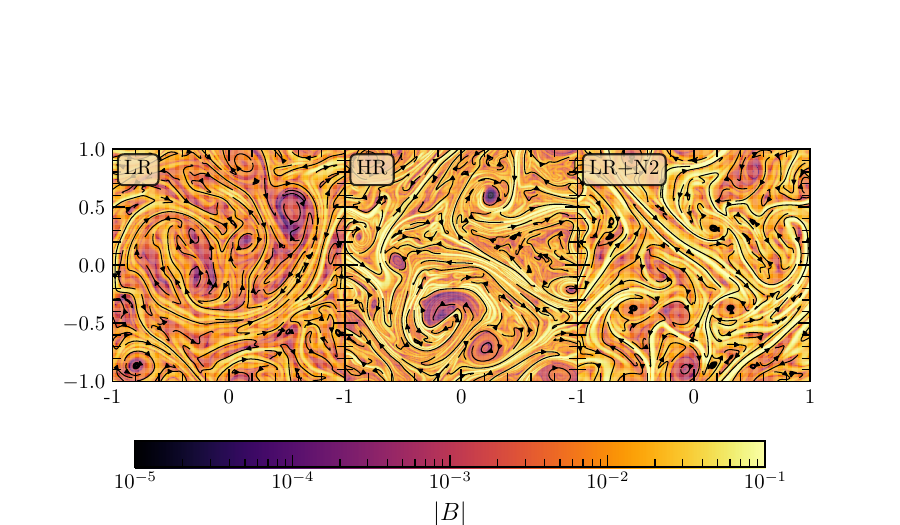}
  \caption{ The magnetic field strength at the end $(t=20)$ of the KHI simulation. Left and central panels denote the fiducial low and high resolution runs respectively, without a subgrid model. Right panel shows low resolution configuration with N2 model activated (hyperparameters in Tab. \ref{tab:newt2dhyp}). Streamlines of the magnetic field are shown in black.}
  \label{fig:newt2db}
\end{figure*}

To validate the a posteriori performance of the models, we perform simulations at LR with the subgrid terms activated, provided by a variety of models. In Fig. \ref{fig:newt2db} we demonstrate the magnetic field strength at the end of the simulation at $t=20$ for our fiducial LR and HR simulations, as well as one representative configuration at LR with a subgrid model activated, model N2. We note that the development of the KHI and associated turbulence is highly chaotic, and so we do not expect to see the low resolution run with subgrid model to exactly match the high resolution picture. We however do see that the amplification of the magnetic field, shown by the colour map in Fig.  \ref{fig:newt2db}, is of a comparable strength between the HR and LR+N2 models, both much stronger than the LR configuration, and that small scale turbulent features have developed in the magnetic field structure. We subsequently demonstrate the similarity between the HR and LR+N models with more  detailed diagnostic quantities.

\begin{figure}[t]
	   \centering 
	     \includegraphics[width=0.49\textwidth]{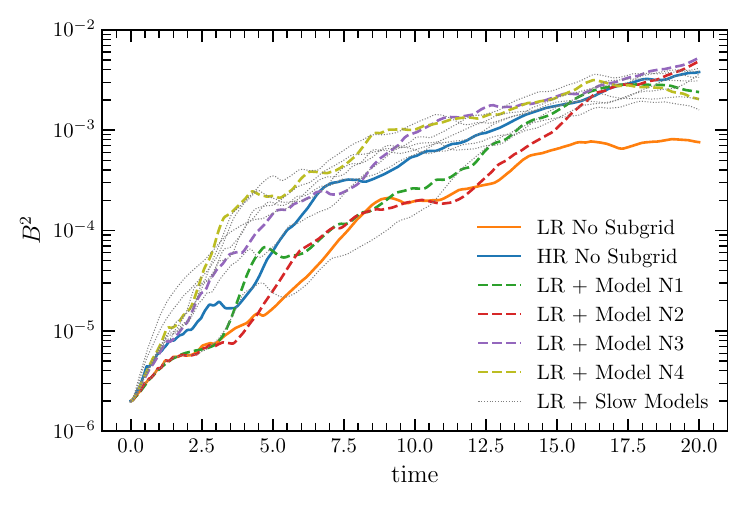}
	     \caption{ Amplification of magnetic field energy with subgrid models. Solid lines represent reference runs without subgrid models, dashed lines represent runs with subgrid models that are faster than run HR. In dotted grey we demonstrate a selection of other models which successfully reproduce the amplification of run HR, but are slower to evaluate.}
	  \label{fig:newt2dbamp}
	 \end{figure}

In Fig. \ref{fig:newt2dbamp} we demonstrate the amplification of the magnetic field due to the KHI for the two reference configurations with no subgrid model, LR and HR, compared with 4 models for different choices of hyperparameter.

From our search of the hyperparameter space we reproduce the four best performing models in the a posteriori test that are faster to execute than the HR configuration. Here we select models based on their ability to reproduce the final magnetic field amplification at $t=20$. In grey we demonstrate a further 7 models that perform well in reproducing the magnetic field amplification, but which are slower to execute than model HR, thus giving no gain in performance. In Tab. \ref{tab:newt2dhyp} we detail the final magnetic field strength reached which consistently  exceeds that of the low resolution run by at least a factor of 2.7 across the models considered, as well as the relative speeds of the simulations. These models are stochastically trained, and so we should expect that different models with different hyperparameters, and even different realisations of models with the same hyperparameters, should perform differently. Nevertheless, we find a range of hyperparameters that successfully enhance the amplification of the magnetic field for resolution LR to a level comparable with that of resolution HR. Further, we find that a non trivial subset of these hyperparameters gives us useful models, that are sufficiently fast to evaluate that the LR run with subgrid model is faster than the corresponding HR run. This confirms that we can successfully use subgrid models from NNs to more efficiently capture the turbulent physics of the KHI than with direct high resolution simulations. 

The speed ups over configuration HR are, for our largest models, fairly small. For this 2D Newtonian problem the cost of evaluating the right hand side of the evolution equations is relatively small, and so the impact of adding the NN evaluation has a large impact on the runtime especially for larger models. We test the impact of various hyperparameters on the runtime of the model and find two major observations. Firstly, as expected, the evaluation of a more complex activation function slows the code down notably. The $\tanh(x)$ activation function (as taken from the standard C++ math library) is consistently $\sim{}1.5\times$ slower to evaluate than the ReLU activation function, and so only the smallest models are feasible with this activation function. Unsurprisingly $N_n$ has the most significant impact on speed of execution, and of the model architectures we consider, it is clear that only models with 48 or fewer neurons per layer can give a faster performance than the HR configuration, with this threshold decreasing as the number of hidden layers increases.

\begin{figure}[t]
	\centering 
	\includegraphics[width=0.49\textwidth]{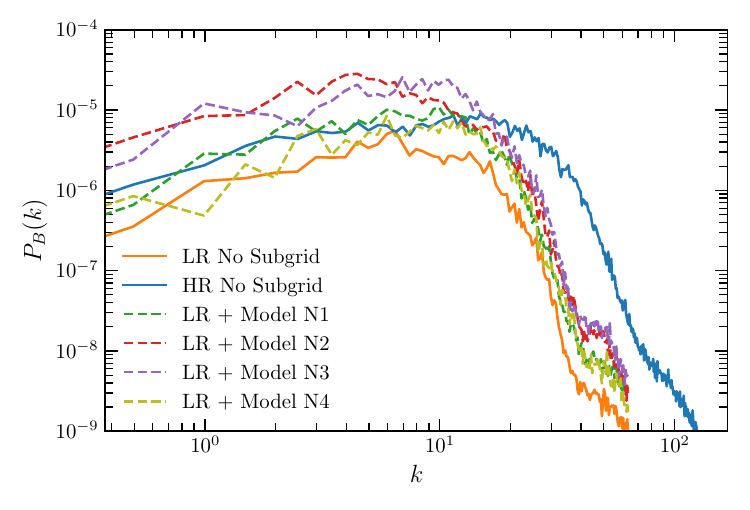}
	\caption{ Magnetic field spectrum at $t=20$ in 2D Newtonian KHI. Solid lines represent reference runs without subgrid models, dashed lines represent runs with subgrid models, detailed in Tab \ref{tab:newt2dhyp}.}
	\label{fig:newt2dspec}
\end{figure}

We demonstrate also the spectrum of the magnetic field that develops by the end of the simulation, in Fig. \ref{fig:newt2dspec}. Clearly, the HR configuration is capable of capturing small scale magnetic field structure that is impossible to resolve in a LR run, no matter the presence of a subgrid model, due to the discretisation length. However, we note, in line with other similar studies \cite{Vigano:2019a}, that the addition of the subgrid terms successfully captures the growth of the magnetic field at longer wavelengths, present in the HR run and in the N1,N2 and N3 subgrid models.

\begin{table*}[t]
	   \centering    
	   \caption{Hyperparameters for models of 2D Newtonian KHI}
	   \begin{tabular}{c|cccccc|cc}        
		     \hline
		     Name & $N_L$ & $N_n$ & Act. Func. & Initial LR &  $t_{\mathrm{start}}$ & $\sigma$ & Speed $t/hr$ on 1 CPU  & $B^2(t = 20)$\\
		     \hline
		     LR &-&-&-&-&-&-&  7.32 & $7.58\times 10^{-4}$ \\
		     HR &-&-&-&-&-&-& 0.964 & $3.78\times 10^{-3}$\\
		     \hline
		     N1 & 2  & 16 & $\tanh$ & $10^{-4}$ & 50 & 0 &  1.42 & $2.39\times 10^{-3}$\\
		     N2 & 2  & 48 & ReLU & $10^{-3}$ & 50 & 0 & 0.994 & $4.86\times 10^{-3}$\\
		     N3 & 2  & 32 & ReLU & $10^{-3}$ & 75 & 10 & 1.28 & $5.25\times 10^{-3}$\\
		     N4 & 4  & 16 & ReLU & $10^{-4}$ & 75 & 10 & 1.34 & $2.03\times 10^{-3}$\\
		     \hline
		   \end{tabular}
	  \label{tab:newt2dhyp}
	 \end{table*}

\begin{figure}[t]
	\centering 
	\includegraphics[width=0.49\textwidth]{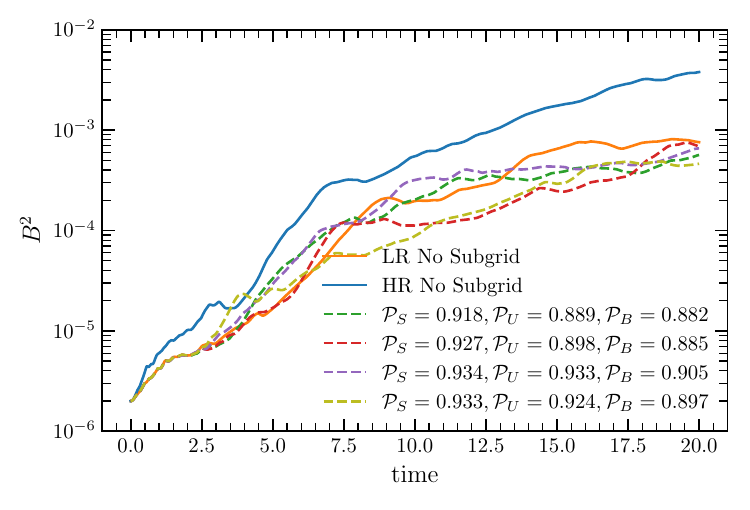}
	\caption{ Amplification of magnetic field energy with subgrid models with good a priori performance, but poor a posteriori performance. Solid lines represent reference runs without subgrid models, dashed lines represent runs with subgrid models. The average Pearson correlation coefficient for the components of each subgrid tensor are shown in the legend.}
	\label{fig:newt2d_apriori_apost_bamp}
\end{figure}

The Pearson correlation coefficients demonstrated above for the models we have considered are less accurate than, for instance, those presented for the models considered in \cite{Rosofsky:2020a}, or for the gradient models considered in \cite{Vigano:2019a}. We emphasise however that our selection criterion for models has been their a posteriori performance, not a priori performance. As an example of the relative importance of both of these tests, we demonstrate a second set of models in Figure \ref{fig:newt2d_apriori_apost_bamp}.

Here we see that for all models considered the amplification of the magnetic field energy is consistent with that of the LR configuration, and the expected amplification of the magnetic field to the HR level is absent. Compared to the models considered above however, the average value of the Pearson correlation coefficient is much higher. A purely a priori selection criterion might select such models, however it is clear that they do not capture the physical effect desired. This failure may be due to the models being well trained on the non-turbulent bulk of the training set, leading to a good performance in the a priori test, where the majority of points have small values of the subgrid tensor, which are accurately predicted; but poor performance in the tail of the dataset, where subgrid tensor values are large. This failure is not identified by the a priori test, but is by the a posteriori test. Since the a priori test is constructed from a dataset with no subgrid model present only the a posteriori test can validate whether the model is capable of accurately evolving the fluid state that incorporates the model's own feedback into the evolution.

\subsubsection{The role of individual subgrid tensors} \label{sec:newt_termbyterm}

In the results above we have looked at modelling all the subgrid terms that arise from non-commutation of the filtering operation with the non linear fluxes in the MHD equations. In contrast, other works constructing such models have focused on the impacts of specific terms. For instance in \cite{Radice:2017zta} only the subgrid term in the momentum equation is incorporated, while in \cite{Rosofsky:2020a} pressure dependent terms are neglected. In the gradient approach of \cite{Vigano:2020ouc} all terms are considered, however in production applications such as \cite{Palenzuela:2021gdo}, only the term in the induction equation is incorporated, since in \cite{Aguilera-Miret:2020dhz}  it is argued that the most important term to include in the evolution to capture magnetic field amplification at the merger of a BNS system is the subgrid term in the induction equation, while other terms can suppress this amplification. We here show the role of each term, one-by-one in the evolution.

\begin{figure}[t]
	\centering 
	\includegraphics[width=0.49\textwidth]{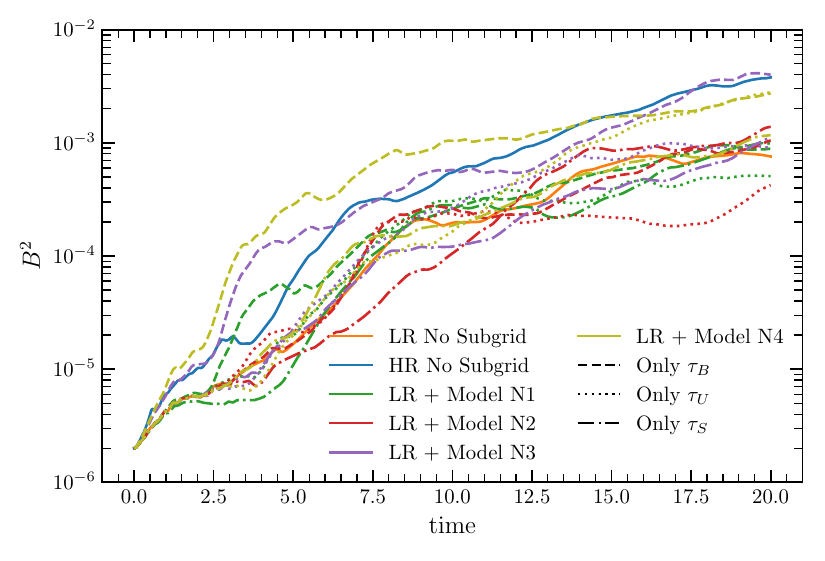}
	\caption{ Amplification of magnetic field energy with subgrid models only active in one evolution equation at a time. Solid lines represent reference runs without subgrid models, dashed lines represent runs with subgrid models only active in the induction equation (Eq. \ref{eq:sgnewtB}), dotted lines represent runs with subgrid models only active in the energy equation (Eq. \ref{eq:sgnewtU}) and dot-dashed lines represent runs with subgrid models only active in the momentum equation (Eq. \ref{eq:sgnewtS})  . Colours represent the different models detailed in Tab. \ref{tab:newt2dhyp}.}
	\label{fig:termbyterm}
\end{figure}

In order to determine the relative importance of these different terms we take our best performing models and run each simulation again, with only one subgrid term present at a time. We show the obtained amplification in Fig. \ref{fig:termbyterm}. We make the following observations. Generally, the magnetic field amplification is largest for configurations with only the subgrid term in the induction equation present (Only $\tau_B$ configurations), as opposed to any other single term, supporting the suggestion of \cite{Aguilera-Miret:2020dhz} that this is the term most important for the amplification. However, we also note that for all models but model N3, it is possible to achieve amplification commensurate with the ``Only $\tau_B$'' evolution by incorporating a different term , for instance, we see for model N4 that the final amplification for the Only $\tau_B$ configuration is approximately identical to the ``Only $\tau_U$'' configuration. It is also clear that for models N1 and N2, that, while the evolution with all subgrid terms active presented in Fig. \ref{fig:newt2dbamp} shows a large amplification consistent with the HR No Subgrid run, none of the evolutions with only one subgrid term presented in Fig. \ref{fig:termbyterm} show an amplification much larger than the LR No Subgrid evolution. From the array of models we consider therefore, it is difficult to conclude that one subgrid term can be prioritised over all other terms, and we suggest that all terms must be included to fully capture the turbulent physics in play. 

A note of caution on our findings. We have presented a set of models picked due to a posteriori criteria, that is, models which, with all subgrid terms included, reproduce the amplification of the HR No Subgrid configuration. We enforce no selection criterion on the individual performances of the 3 NNs that comprise the overall model. It is therefore possible that, for any given model N?, the error budget is unevenly spread amongst the 3 networks, and only the combination of all networks gives an accurate result. However, when analysing the errors demonstrated in Fig. \ref{fig:newt2d_training}, we can see that, while the magnitude of errors may differ between the models, the hierarchy of errors between the three networks is identical between the models, suggesting the partitioning of errors may be relatively consistent. We also note that model N3 has the lowest training error by the end of the training process. These facts lend slightly more robustness to the idea that $\tau_B$ may be the more important subgrid term, though again, it seems difficult to conclude that ignoring the other terms does not miss key physics, especially in the transport of angular momentum. 

\subsection{Special Relativistic Kelvin-Helmholtz Instability in 3D}

We now repeat the above analysis generalising to 3D, and moving from Newtonian to SR physics. 

\subsubsection{Training Dataset construction}

We evolve the initial data  detailed in Sec. \ref{sec:SR_ID} using the code \GRAthena , up to $t=40$ at resolutions LR, SR and HR, corresponding to 128, 256 and 512 points in each direction over the domain. As in Sec. \ref{sec:res:newt}, we take the high resolution data, calculate the subgrid terms appearing in Eqs \ref{eq:sr_sg_1} - \ref{eq:sr_sg_4}, and filter this down to a lower resolution, along with the primitive variables and their spatial derivatives. These will respectively be used as the outputs and inputs of the NNs. We train using the same loss function, stopping criteria and maximum number of epochs as in the Newtonian case. Here we explore two types of model, based on different filtering factors. Models denoted \ft are constructed by treating the SR data as our continuum solution where the subgrid tensors are calculated, and then filtering this data down by a factor of 2 to the LR resolution, where the models are trained. Models denoted \ff are constructed by filtering instead the HR data down by a factor of 4 to the LR resolution. Practically this filtering operation means that a neighbouring cube of 8 (64) computational cells are averaged to a single cell for constructing the training dataset for the \ft (\ff) models. In our a posteriori tests the \ft models should reproduce the behaviour of the SR simulation, while the \ff models should reproduce that of the HR simulation.

We note a key difference between the Newtonian and special relativistic problems we consider, that the first is 2D and the latter 3D. Since a given timeslice of data from our simulations is much larger in terms of computational cells in 3D, we have much more training data available to us, and therefore our dataset construction and data augmentation strategies are different compared to the 2D Newtonian case. As in the Newtonian case, we construct our training set from timeslices taken after an initial burn in time $t_{\mathrm{start}}$, in order to train on purely turbulent data, rather than data that carries the imprint of the initial conditions. Since the models we train are relatively small in terms of trainable parameters, we then reduce the datasets further, by taking a spatial subsample of each timeslice in our training simulation after $t_{\mathrm{start}}$, by selecting $N_{\mathrm{sample}}$ \MeshBlocks ~at random from the full 3D data at each timeslice. We refer the reader to \cite{Stone:2020,Daszuta:2021ecf} for a detailed description of the grid structure of \GRAthena, but note that a \MeshBlock ~ is a spatially contiguous, logically rectangular spatial subdomain of the overall computational grid, which is, for all problems considered in this paper, $16^3$ grid cells in size. We find that the size of these datasets removes the need to augment the data sets as we found in the 2D case. As before, we explore hyperparameter space, as described in App. \ref{app:hyp}, training models that we will subsequently select based on their a posteriori performance. 

\subsubsection{Model training and a priori testing}

\begin{figure}[t]
 	\centering 
 	\includegraphics[width=0.49\textwidth]{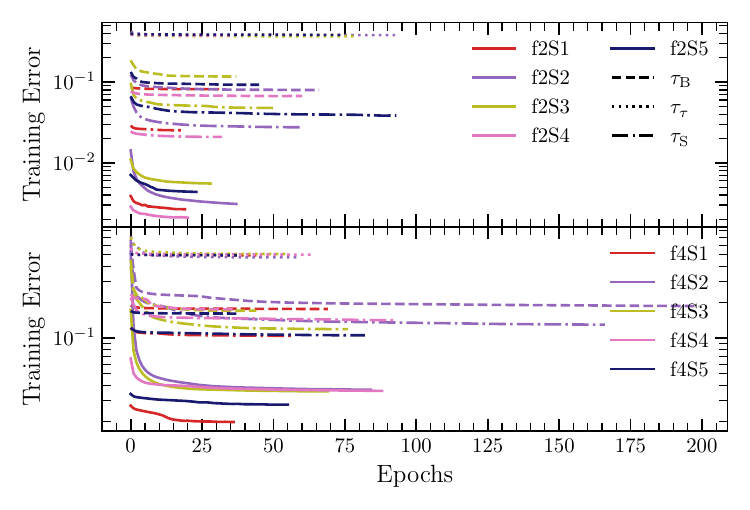}
 	\caption{ Online training error given by mean squared error loss function on training dataset. Colours represent different models, while linestyles represent models trained to reproduce different subgrid terms. Upper panel: \ft models. Lower panel: \ff models.}
 	\label{fig:sr3d_training}
 \end{figure}

Following the same training procedure as in the Newtonian case we monitor the online training error with a validation dataset separated from our training dataset. We demonstrate the model training error in Fig. \ref{fig:sr3d_training}, for a subset of successful models (according to the same a posteriori criterion as before) that we focus our analysis on. The hyperparameters of these models are presented in Tab. \ref{tab:sr3dhyp}. 

We note that the training error in the energy term $\tau_\tau$, is consistently larger than that of the magnetic field, momentum and density terms. Nevertheless, we find that in our a posteriori tests all models show an excellent conservation of energy, at least to a factor of one part in $10^3$, suggesting that this limitation to our model is not a significant source of error in our simulations. \ft models also show a lower error than their \ff counterparts, in agreement with their overall performance in the a priori test.

\begin{figure*}[t]
  \centering 
  \includegraphics[width=0.32\textwidth]{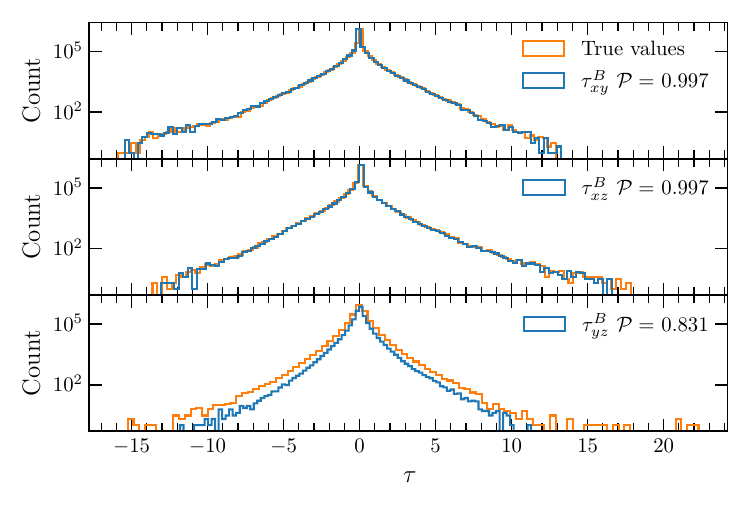}
  \includegraphics[width=0.32\textwidth]{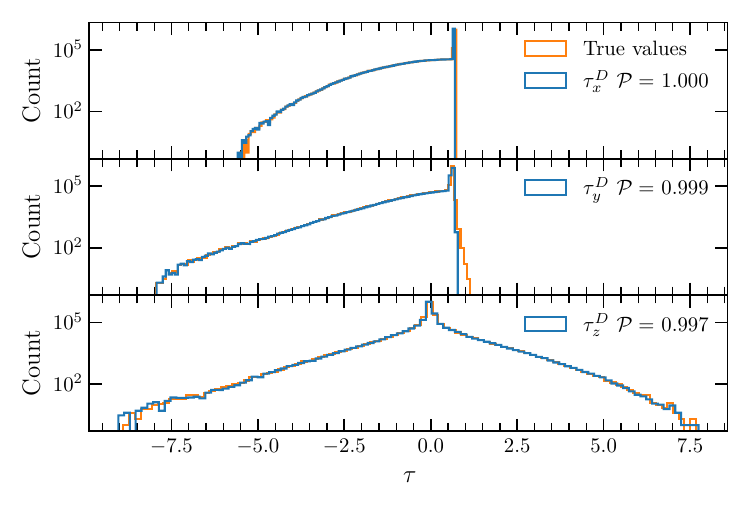}
  \includegraphics[width=0.32\textwidth]{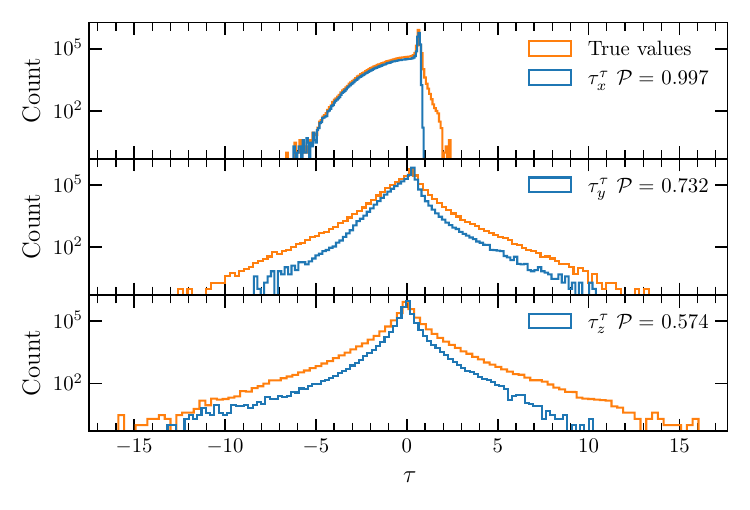}\\
  \includegraphics[width=0.99\textwidth]{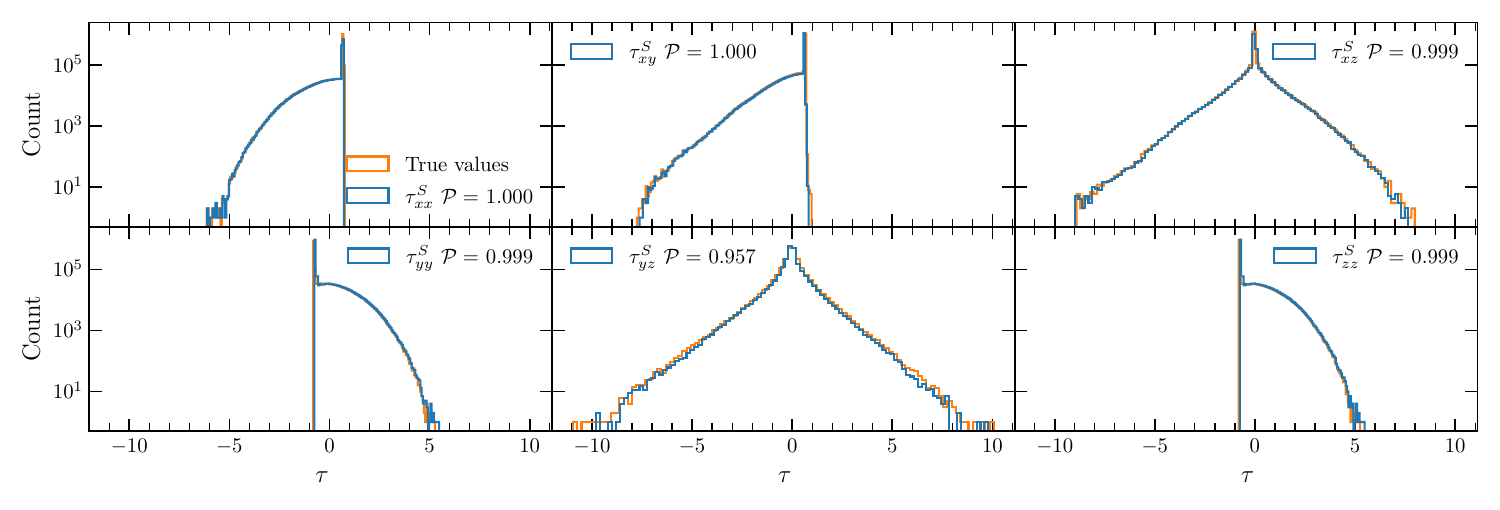}
  \caption{Histogram of the subgrid tensors in the testing dataset (orange) and the predicted values of those tensors by the NN model \texttt{f2S1}. On the x-axis we show the value of the normalised subgrid tensor. On the y-axis we show the unnormalised count of points in the bins. Left, middle and right upper panels show the subgrid tensors $\tau_B,\tau_D, \tau_\tau$ respectively, while the lower panel shows $\tau_S$. Subpanels show different components of the tensor, indicated in the legend. In the legend we display the Pearson correlation coefficient between the two datasets. } 
  \label{fig:sr3d_apriori_histo_f2}
\end{figure*}

\begin{figure*}[t]
  \centering 
  \includegraphics[width=0.32\textwidth]{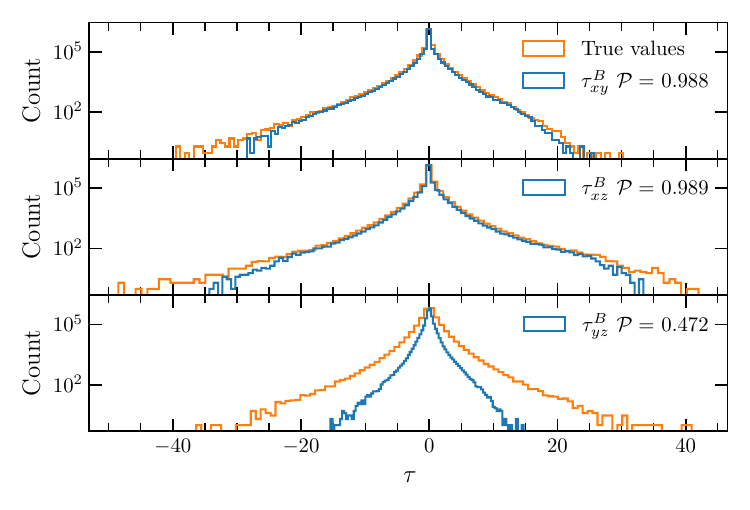}
  \includegraphics[width=0.32\textwidth]{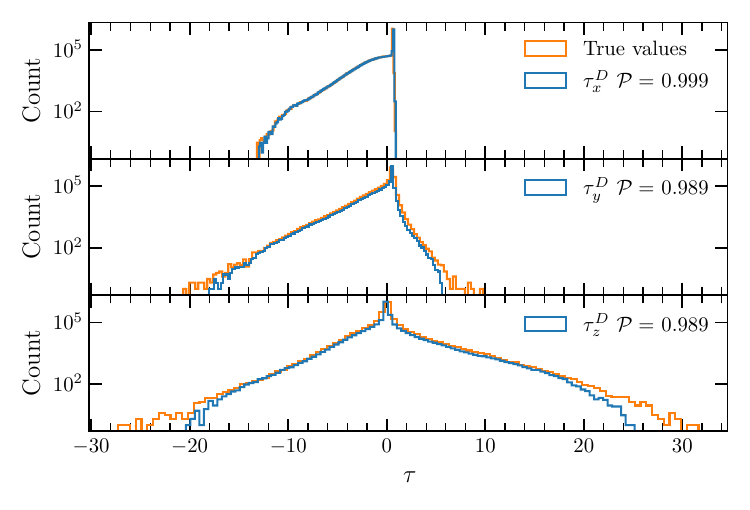}
  \includegraphics[width=0.32\textwidth]{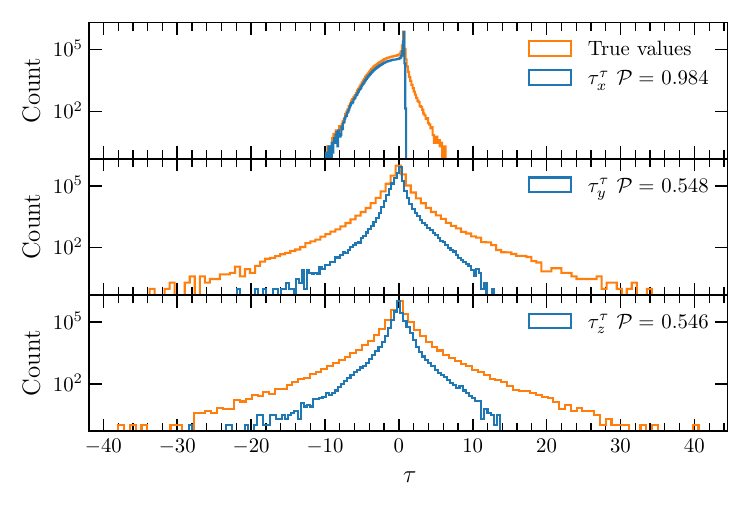}\\
  \includegraphics[width=0.99\textwidth]{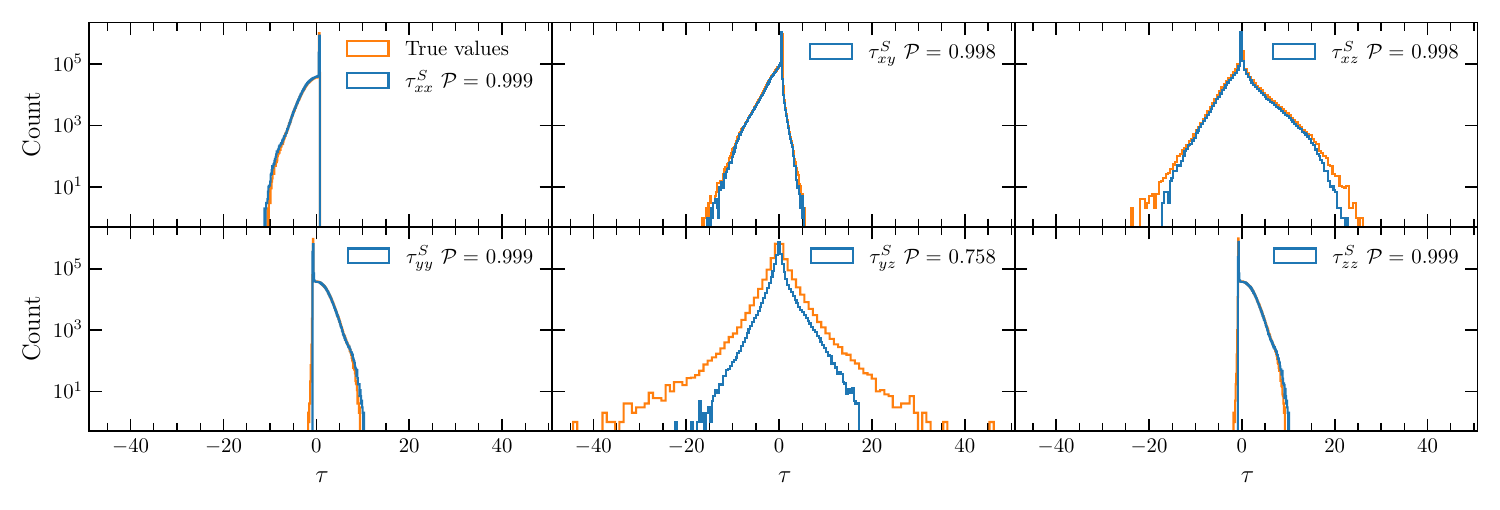}
  \caption{Histogram of the subgrid tensors in the testing dataset (orange) and the predicted values of those tensors by the NN model \texttt{f4S1}. On the x-axis we show the value of the normalised subgrid tensor. On the y-axis we show the unnormalised count of points in the bins. Left, middle and right upper panels show the subgrid tensors $\tau_B,\tau_D, \tau_\tau$ respectively, while the lower panel shows $\tau_S$. Subpanels show different components of the tensor, indicated in the legend. In the legend we display the Pearson correlation coefficient between the two datasets. } 
  \label{fig:sr3d_apriori_histo_f4}
\end{figure*}

We demonstrate also the a priori performance of two of the trained models on testing datasets in histograms in Figs. \ref{fig:sr3d_apriori_histo_f2}, \ref{fig:sr3d_apriori_histo_f4}. As above, we take a set of testing data that is independent of the training dataset, taken from a separate timeslice within the same simulation, and demonstrate the ability of the model to accurately infer the subgrid term. 

As in the Newtonian case, while we only show one case for each set of models \ft and \ff, the other models within these families show similar behaviour. Firstly we note that, in comparison to the Newtonian case, many Pearson coefficients are much higher, in excess of 0.99, with all models other than $\tau_\tau$ exceeding 0.83. We attribute this to the different nature of the training dataset, with the large dataset no longer requiring a suppression of the mean and enhancement of the tails as detailed in App. \ref{app:hyp}. As in the Newtonian case, the worst performing components are associated to the coordinate directions in which there are the smallest overall perturbations in the initial data, especially the $\tau_\tau^z$. As noted above, the worse performance of this term does not impact energy conservation.  In comparison to the \ft models, the \ff models perform notably worse in the a priori test. This is consistent with results seen in the gradient model and for other NN implementations \citep{Vigano:2019a,Rosofsky:2020a}, in which the larger the filtering factor used, the worse the model performs in a priori tests. Since we select our models based on the a posteriori performance, these models are therefore not necessarily the best performers in the a priori test.

We note that the performance of the model is broadly consistent with similar plots produced in \cite{Rosofsky:2020a}, who see correlation coefficients consistently in excess of $\sim 0.95$ for a filtering factor of 2 and in excess of $\sim 0.85$ for a factor 4, though our worst performing components fall significantly below this threshold, a point which holds also for the earlier Newtonian models.

 We attribute this worse performance to two reasons. Firstly, in \cite{Rosofsky:2020a} inputs for a particular model are given by not only the primitives and their first derivatives, as employed here, but also the second derivatives, as well as all of these variables on a stencil of neighbouring points. Taking this approach would increase the size of the input layer by a factor of 23, making the inference of the NN much more expensive. This would also require higher derivative terms and more intricate interpolation stencils to reconstruct these derivatives to the cell faces for flux construction within the numerical evolution code. In addition several terms are neglected in the subgrid tensors constructed in \cite{Rosofsky:2020a}, for instance all terms with a pressure dependence. 
 
 Here we prioritise generating small models, capturing all of the relevant physics, and tolerate the associated error, in order to achieve performant evolutions that can capture the high resolution physics more efficiently than by performing the higher resolution DNS. We note however that our current simulations are in special relativity. In future applications, in general relativity, a given timestep of our code will be much more expensive due to the requirement to solve the Einstein equations. This extra cost will mask the cost of the NN evaluation much more, allowing us to train much larger models, without risking making the code too inefficient. In this scenario it may be that using a wider stencil and higher derivatives gives a valuable accuracy improvement with minimal performance impact. Again, however we re-emphasise the point of Fig. \ref{fig:newt2d_apriori_apost_bamp}, that merely increasing the a priori correlation is not a guarantee of successful performance in the a posteriori test.

 \begin{table*}[t]
   \centering    
   \caption{Hyperparameters for models of 3D special relativistic MHD KHI. Models labelled \ft are trained on SR data, and \ff on HR data. Speed up to target is measured against SR for \ft models and against HR for \ff models. }
   \begin{tabular}{c|cccccc|cccc}        
     \hline
     Name & $N_L$ & $N_n$ & Act. Func. & Initial LR &  $t_{\mathrm{start}}$ & $N_\mathrm{sample}$ & Speed on 1 CPU $(t / \mathrm{hr})$ & $B^2(t=20)$ & Speed up to target & $ N_{\mathrm{mesh}, \mathrm{eff}}$ \\
     \hline
     LR &-&-&-&-&-&-& $3.80 \times 10^{-2}$ & $2.43 \times 10^{-4}$ & - & 128 \\
     SR &-&-&-&-&-&-& $0.333 \times 10^{-2}$  & $1.48 \times 10^{-3}$ & - & 256\\
     HR &-&-&-&-&-&-& $0.0296 \times 10^{-2}$  & $4.31 \times 10^{-3}$ & - & 512 \\ \hline
     \texttt{f2S1} & 2  & 32 & ReLU & $10^{-3}$ & 30 & 1000 & $1.39 \times 10^{-2}$ & $1.88 \times 10^{-3}$ & 4.17 & 270\\
     \texttt{f2S2} & 2  & 64 & ReLU & $10^{-3}$ & 35 & 200 & $0.931 \times 10^{-2}$ & $1.75 \times 10^{-3}$ & 2.80 & 288\\
     \texttt{f2S3} & 3  & 16 & ReLU & $10^{-3}$ & 20 & 50 & $1.54 \times 10^{-2}$ & $1.62 \times 10^{-3}$ & 4.63 & 159\\
     \texttt{f2S4} & 4  & 16 & ReLU & $10^{-3}$ & 20 & 100 & $1.34 \times 10^{-2}$ & $1.26 \times 10^{-3}$ & 4.02 & 175  \\  				
     \texttt{f2S5} & 4  & 16 & ReLU & $10^{-3}$ & 30 & 200 & $1.31 \times 10^{-2}$ & $1.52 \times 10^{-3}$ & 3.94 & 285\\ \hline
     \texttt{f4S1} & 2  & 32 & ReLU & $10^{-3}$ & 20 & 2000 & $1.20 \times 10^{-2}$ & $4.02 \times 10^{-3}$ & 40.5 & 508\\
     \texttt{f4S2} & 3  & 16 & ReLU & $10^{-3}$ & 35 & 100 & $1.29 \times 10^{-2}$ & $4.75 \times 10^{-3}$ & 43.5 & 541 \\
     \texttt{f4S3} & 3  & 32 & ReLU & $10^{-3}$ & 35 & 100 & $0.986 \times 10^{-2}$ & $3.81 \times 10^{-3}$ & 33.3 & 528\\
     \texttt{f4S4} & 4  & 16 & ReLU & $10^{-3}$ & 35 & 500 & $1.14 \times 10^{-2}$ & $4.46 \times 10^{-3}$ & 38.6 & 428\\
     \texttt{f4S5} & 4  & 32 & ReLU & $10^{-3}$ & 30 & 50 & $0.840 \times 10^{-2}$ & $5.46 \times 10^{-3}$ & 28.4 & 409 \\
     \hline
   \end{tabular}
   \label{tab:sr3dhyp}
 \end{table*}
 
\subsubsection{A posteriori testing}
 
We perform simulations at resolution LR with the trained models from both the \ft and \ff families. We then select the best performing models from our a posteriori test in terms of their ability to reproduce the final magnetic field amplification of either the SR or HR no subgrid run. The best performing models and their hyperparameters are listed in Table \ref{tab:sr3dhyp}. 

\begin{figure*}[t]
  \centering 
    \includegraphics[width=0.19\textwidth]{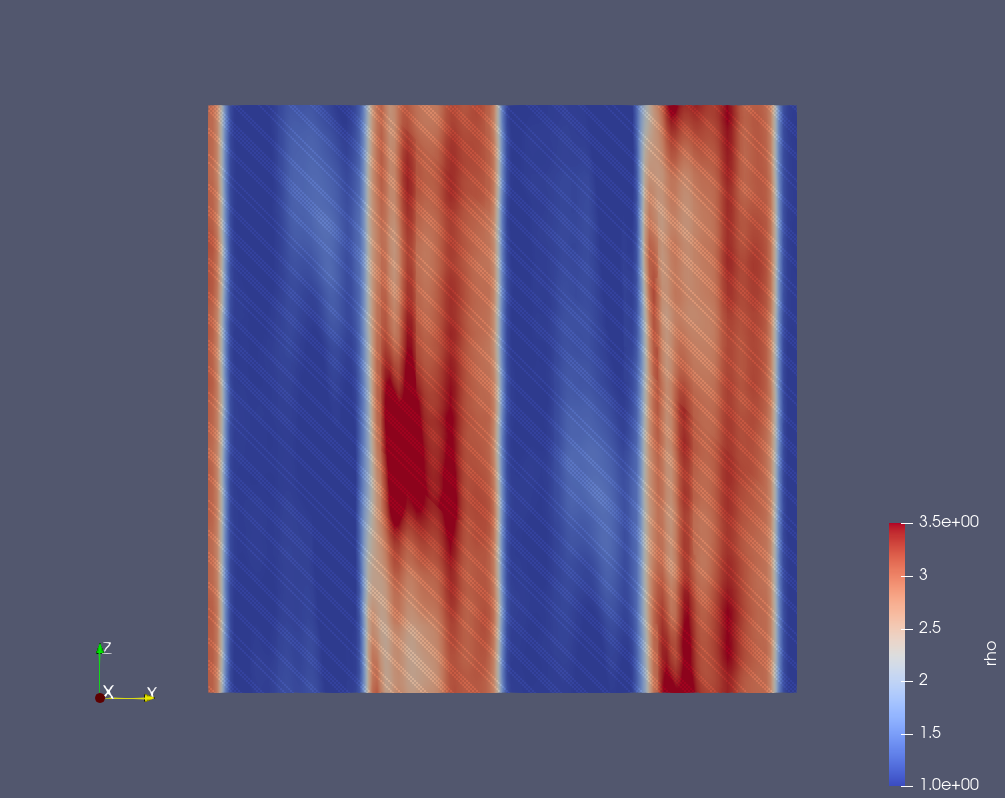} 
\includegraphics[width=0.19\textwidth]{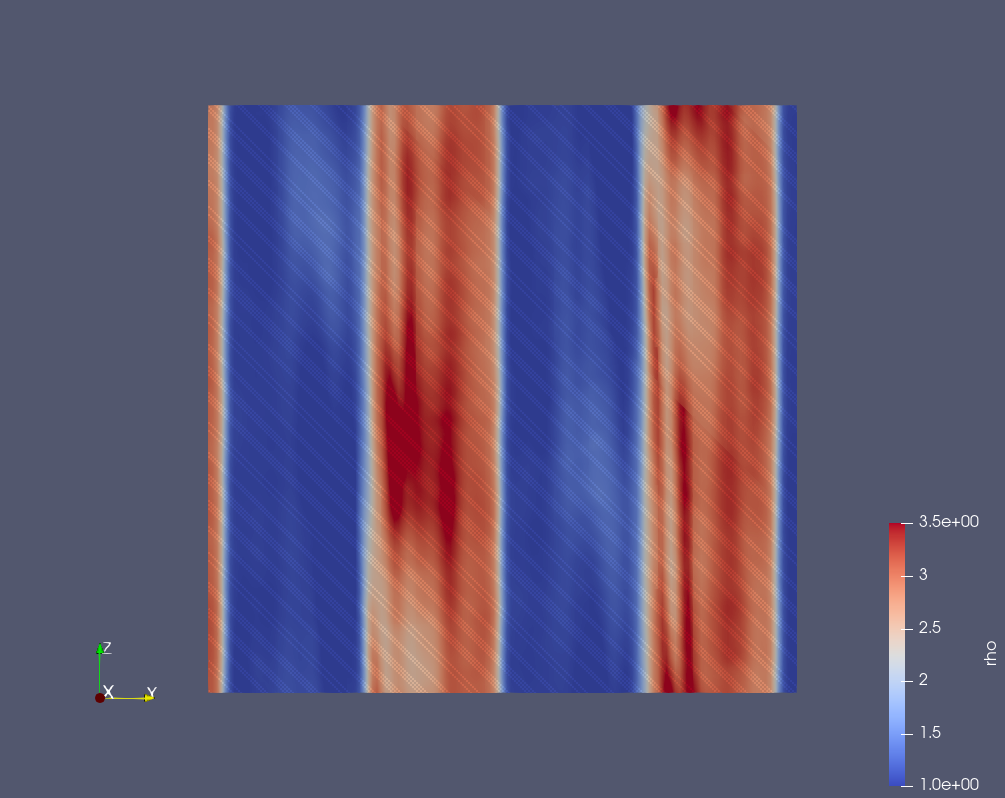} 
\includegraphics[width=0.19\textwidth]{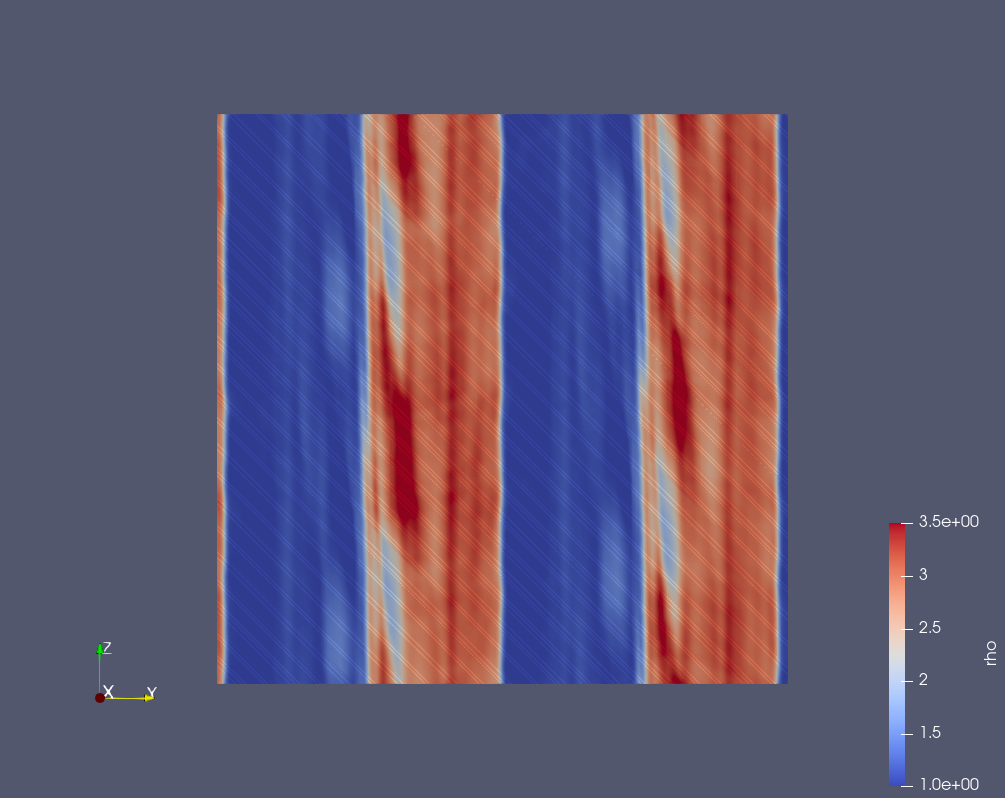} 
\includegraphics[width=0.19\textwidth]{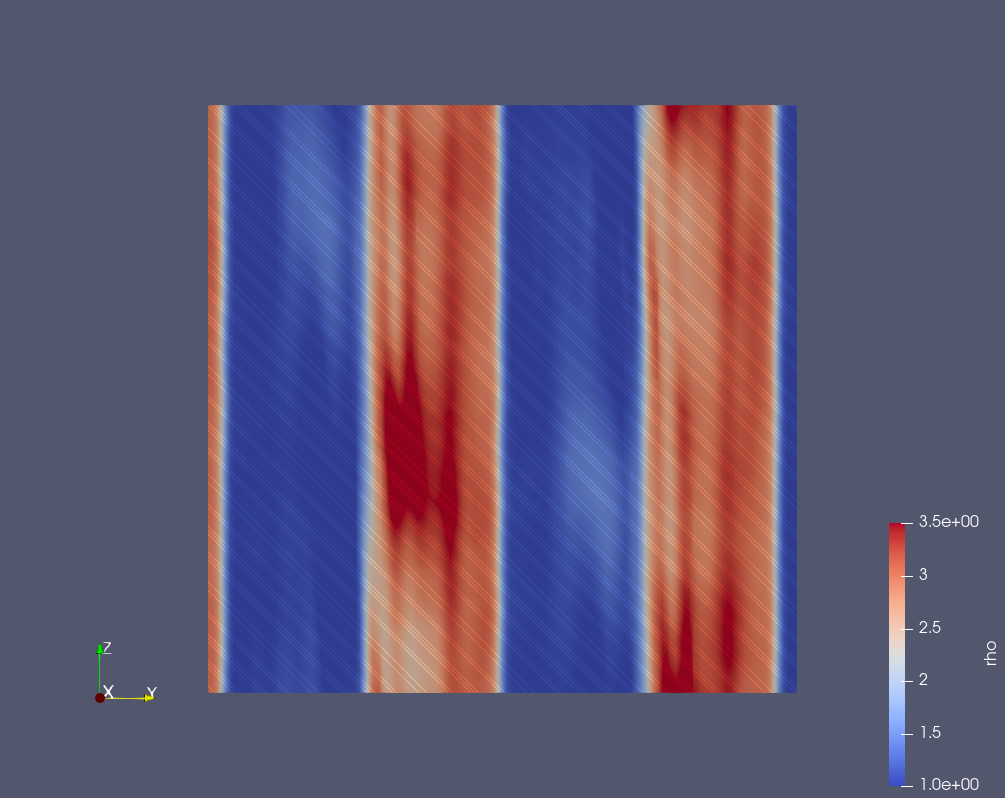} 
\includegraphics[width=0.19\textwidth]{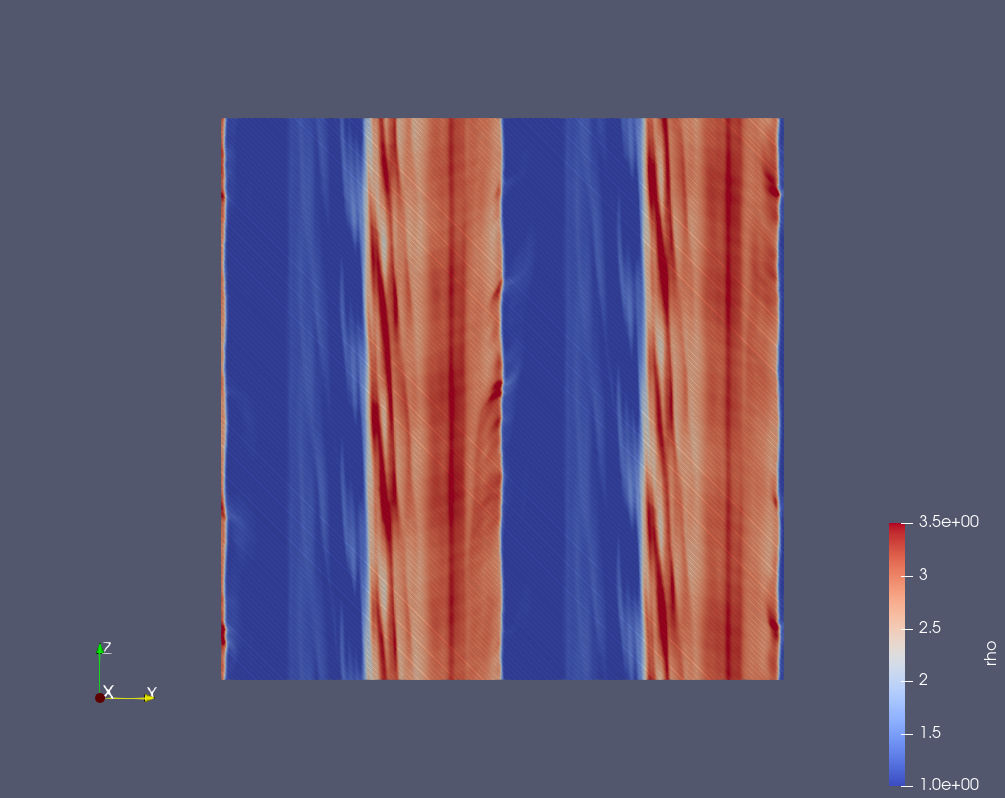}\\
    
        \includegraphics[width=0.19\textwidth]{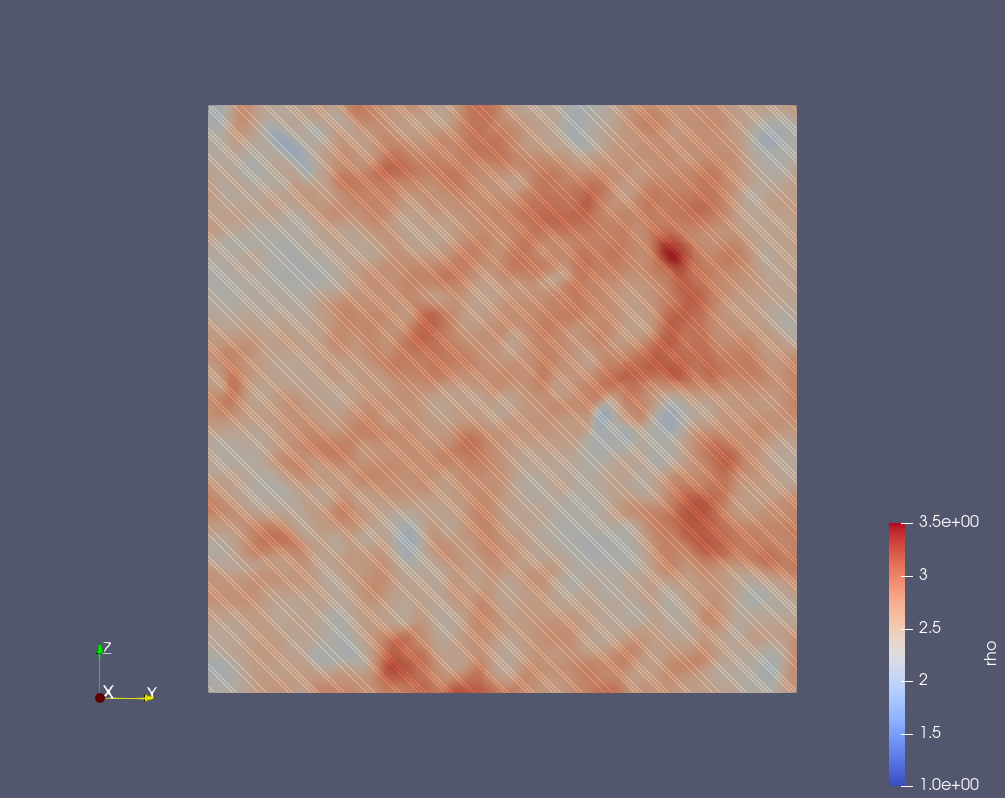} 
\includegraphics[width=0.19\textwidth]{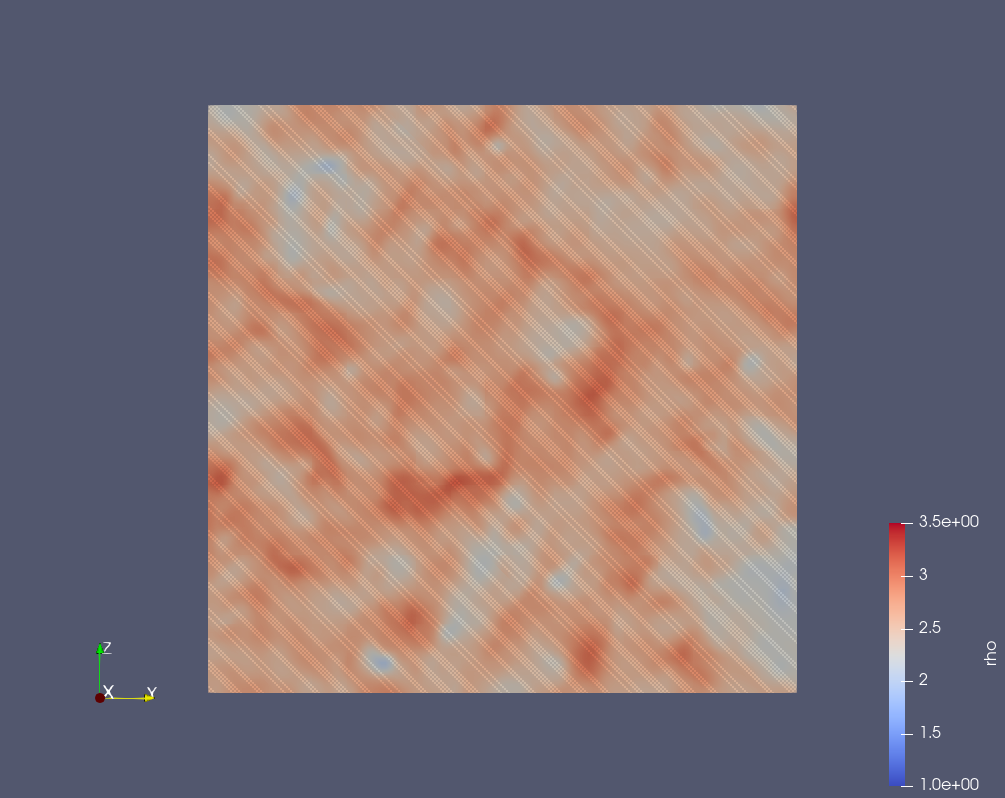} 
\includegraphics[width=0.19\textwidth]{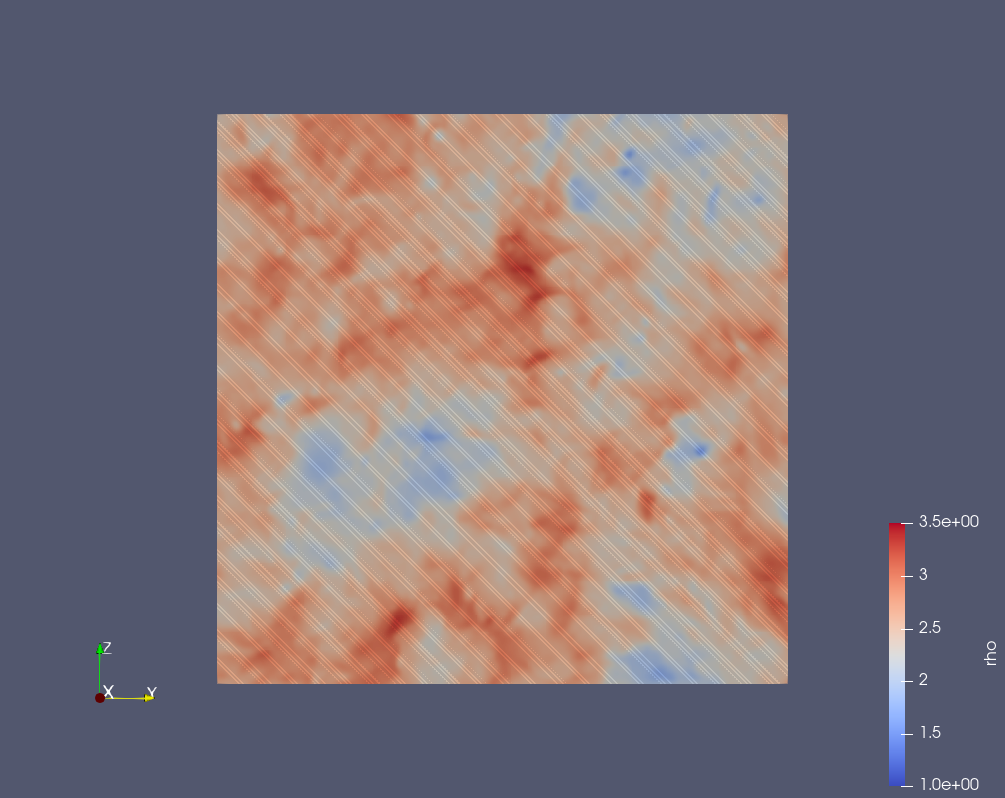} 
\includegraphics[width=0.19\textwidth]{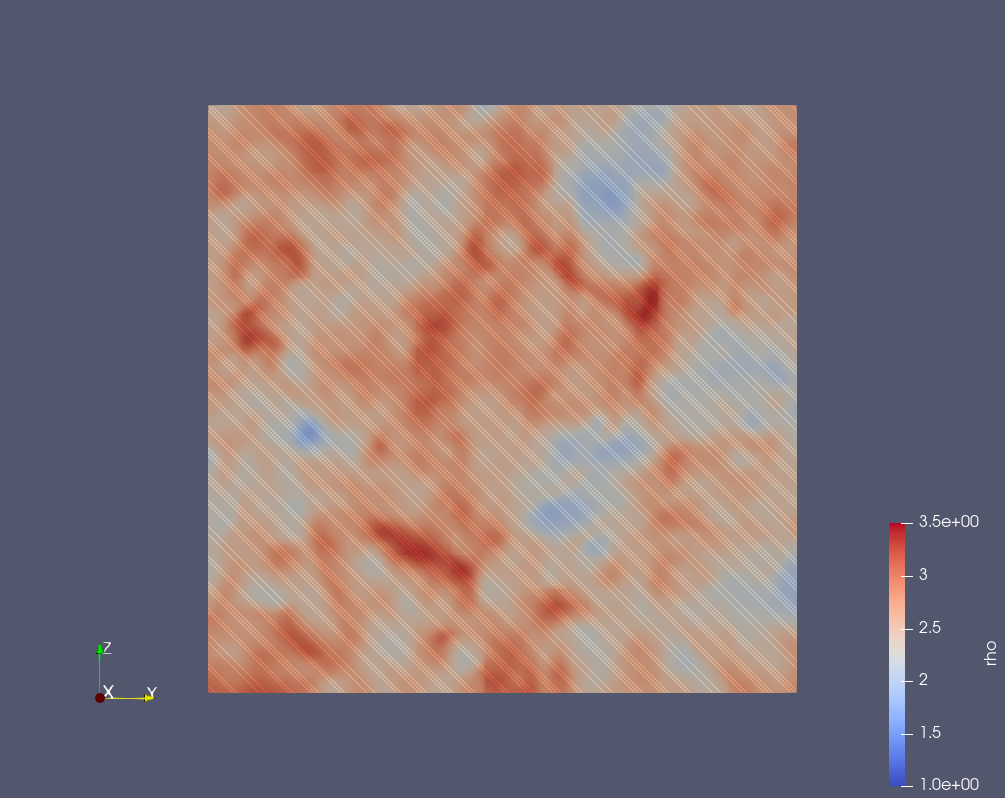} 
\includegraphics[width=0.19\textwidth]{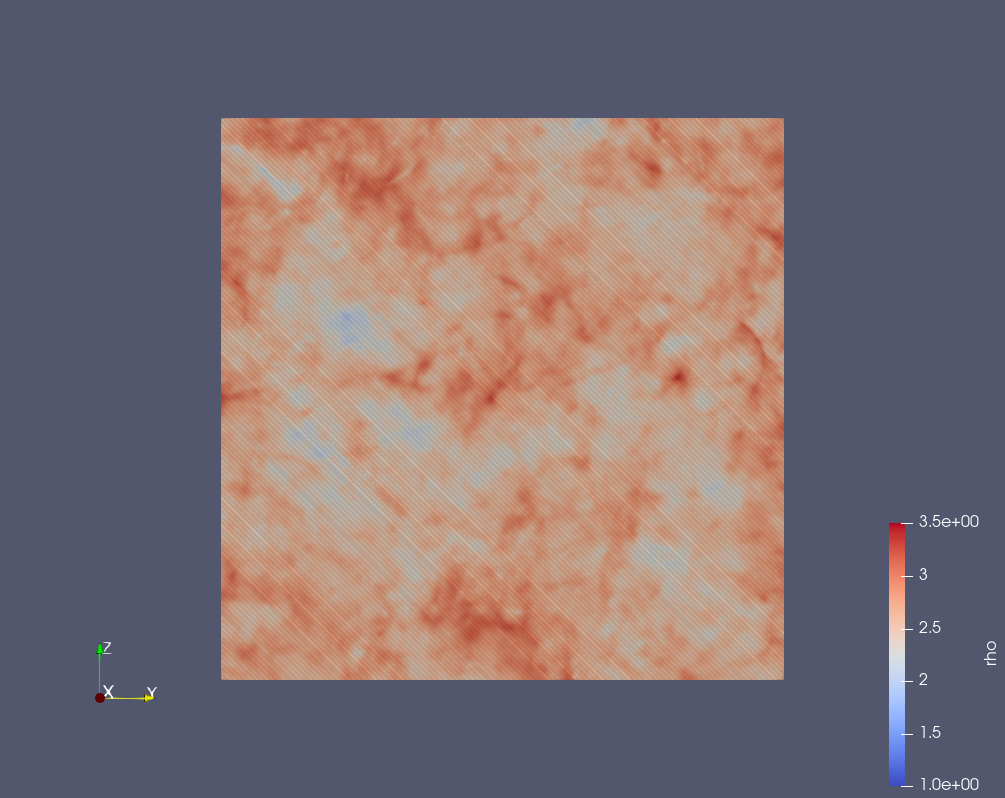}\\
          \includegraphics[width=0.19\textwidth]{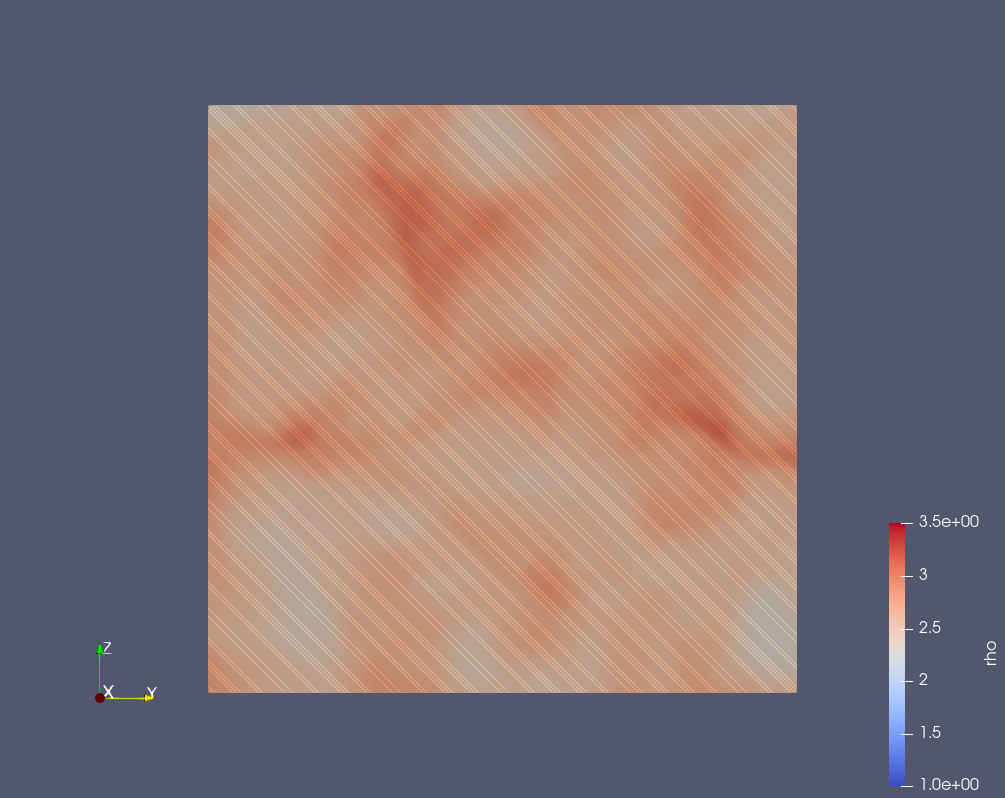} 
\includegraphics[width=0.19\textwidth]{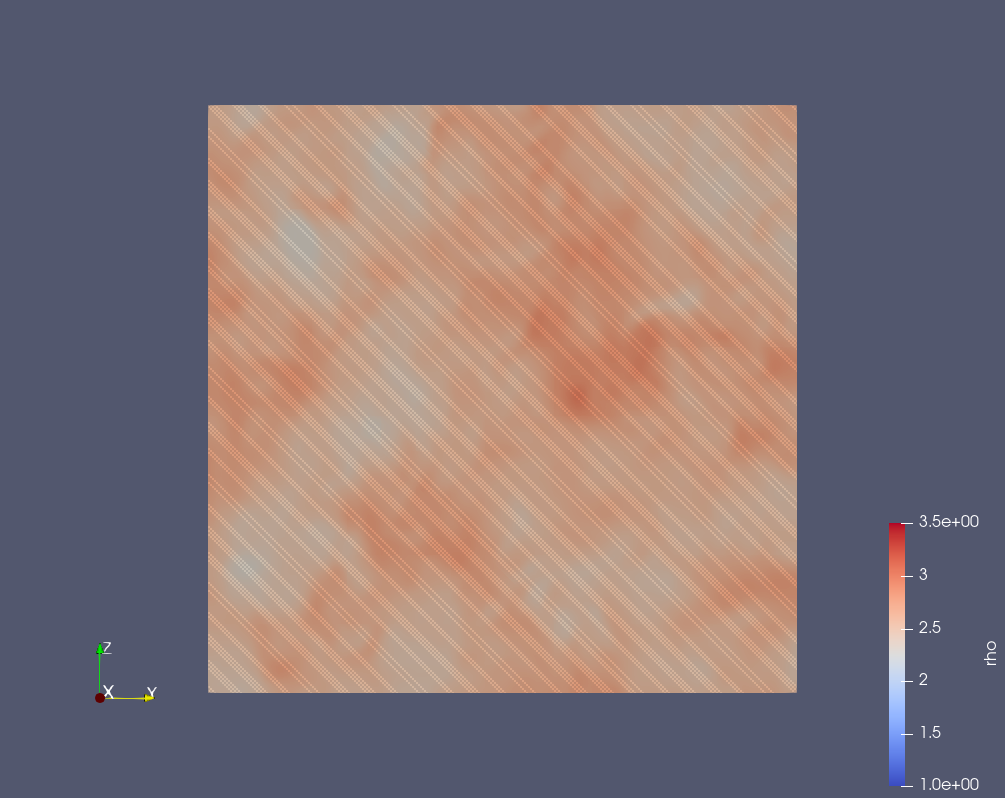} 
\includegraphics[width=0.19\textwidth]{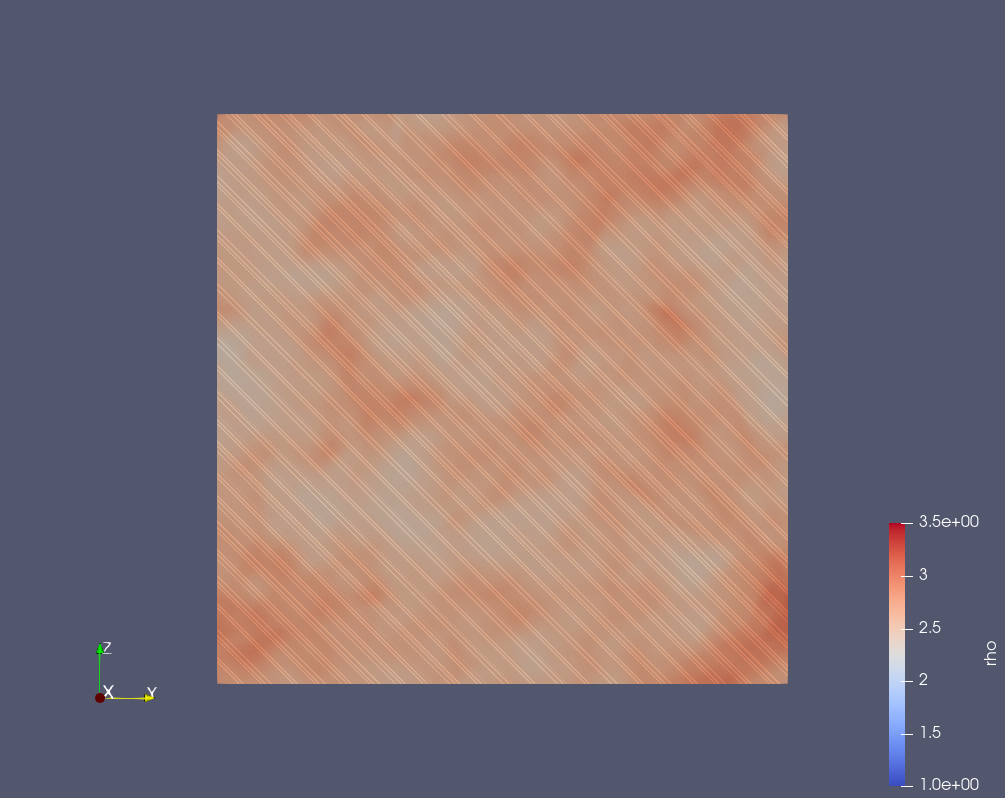} 
\includegraphics[width=0.19\textwidth]{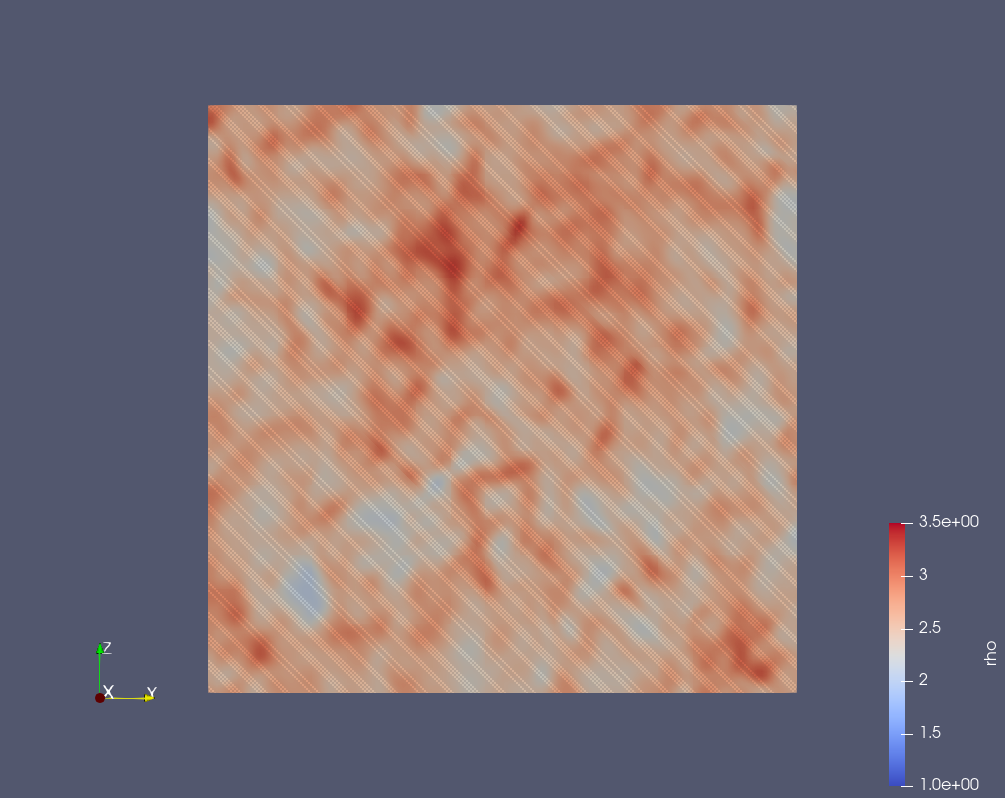} 
\includegraphics[width=0.19\textwidth]{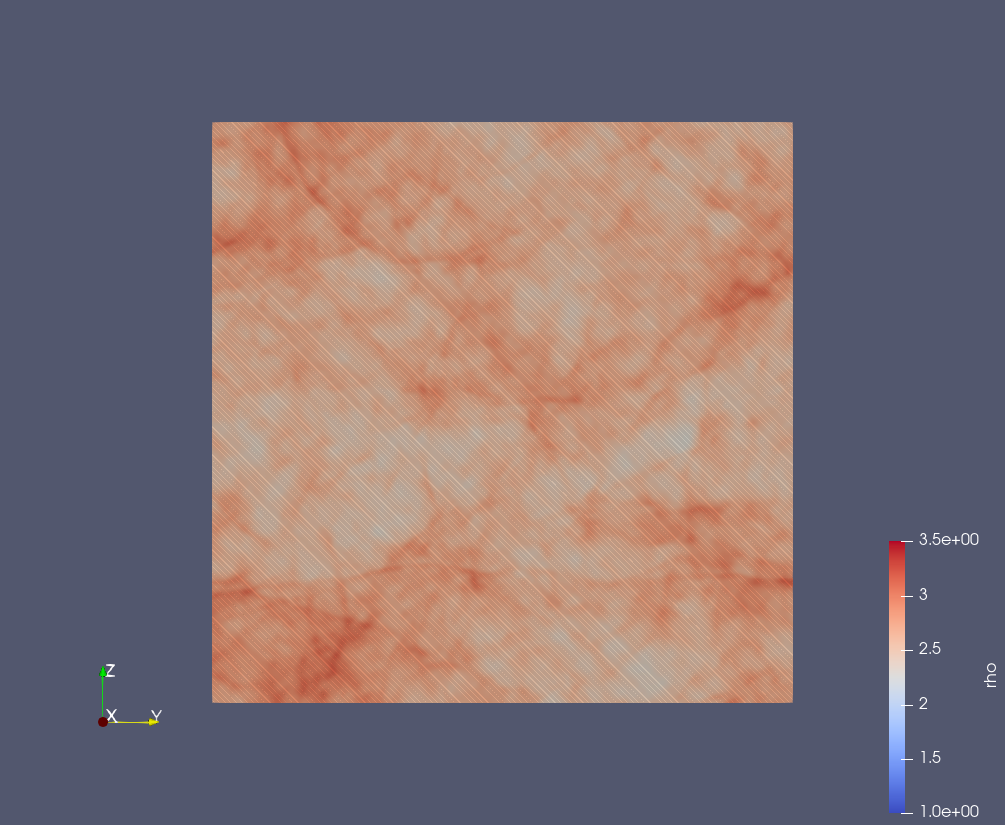}  	
  \caption{ 2D slices of 3D evolution of $\rho$ for special relativistic KHI evolution. The three rows demonstrate different timeslices $t = (0.25,10,20)$. The five columns show the models LR, \texttt{f2S1}, SR, \texttt{f4S1}, HR respectively. }
  \label{fig:sr3d_rho}
\end{figure*}

\begin{figure*}[t]
	\centering 
	\includegraphics[width=0.19\textwidth]{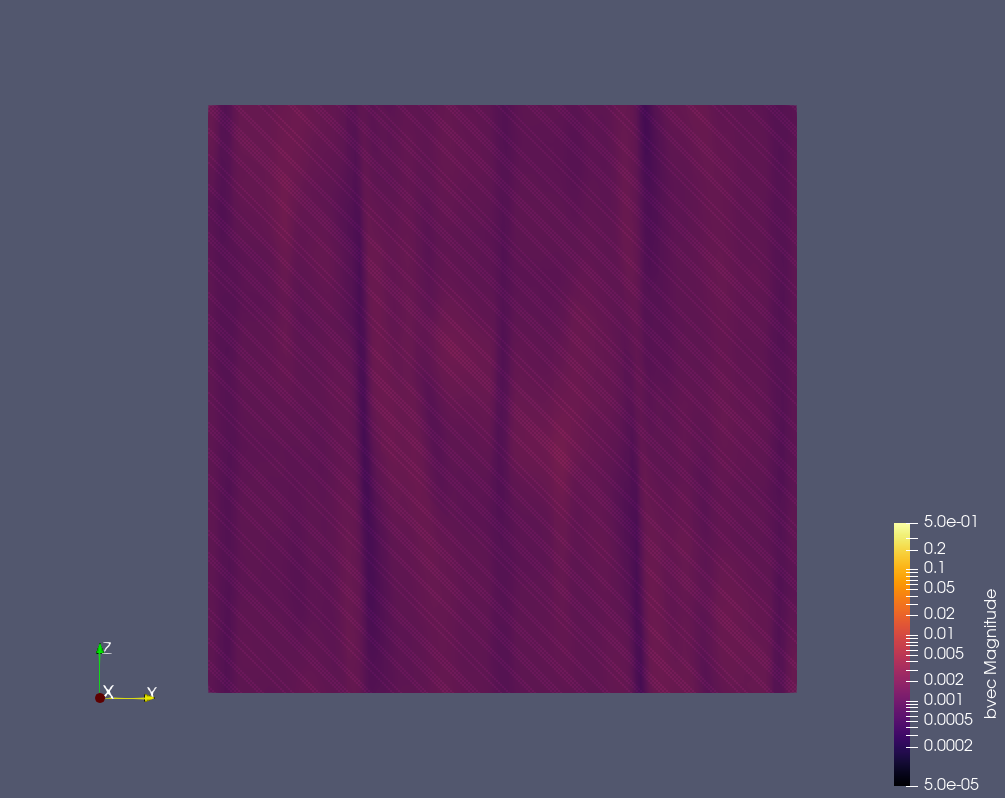} 
\includegraphics[width=0.19\textwidth]{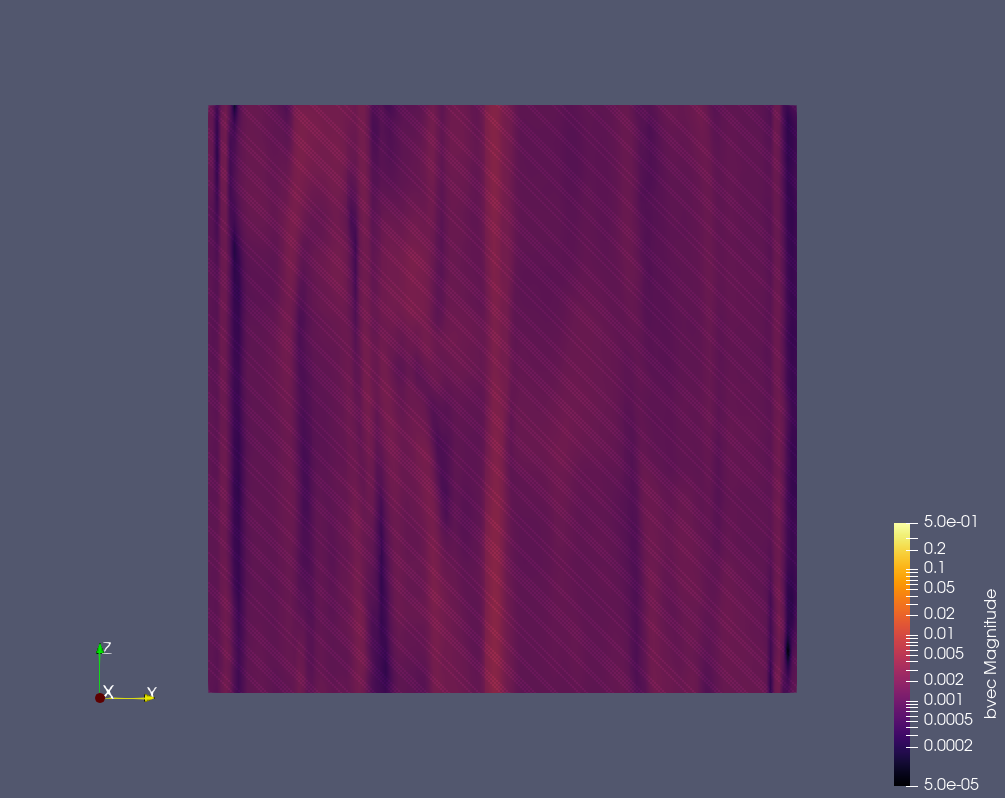} 
\includegraphics[width=0.19\textwidth]{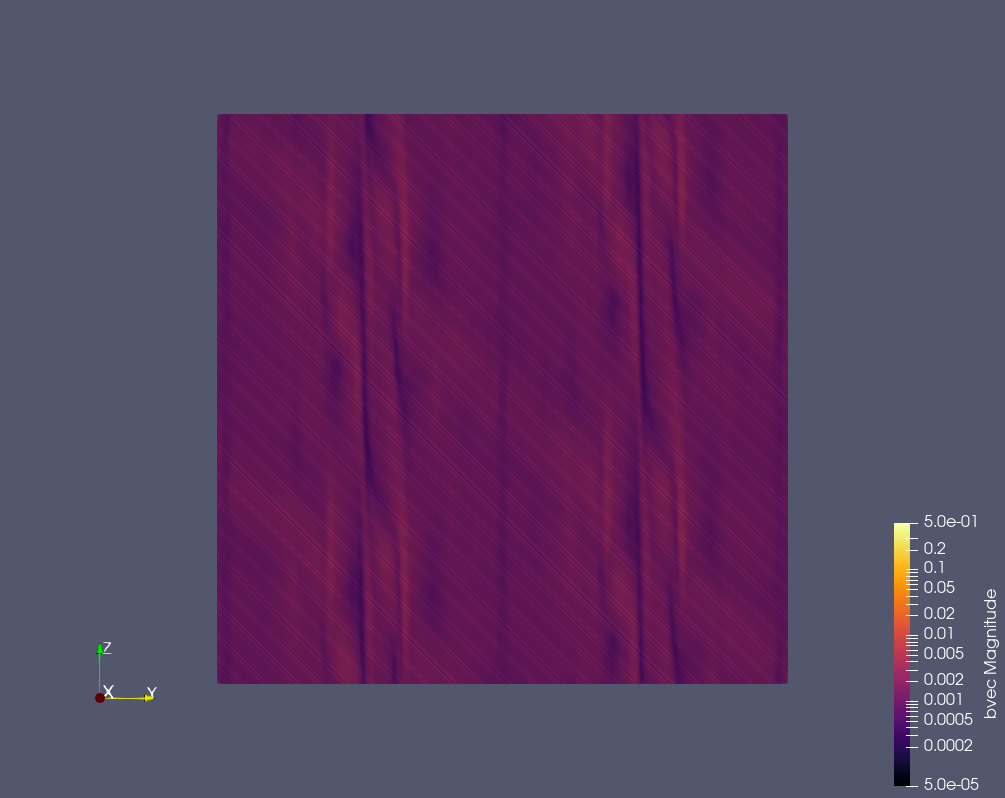} 
\includegraphics[width=0.19\textwidth]{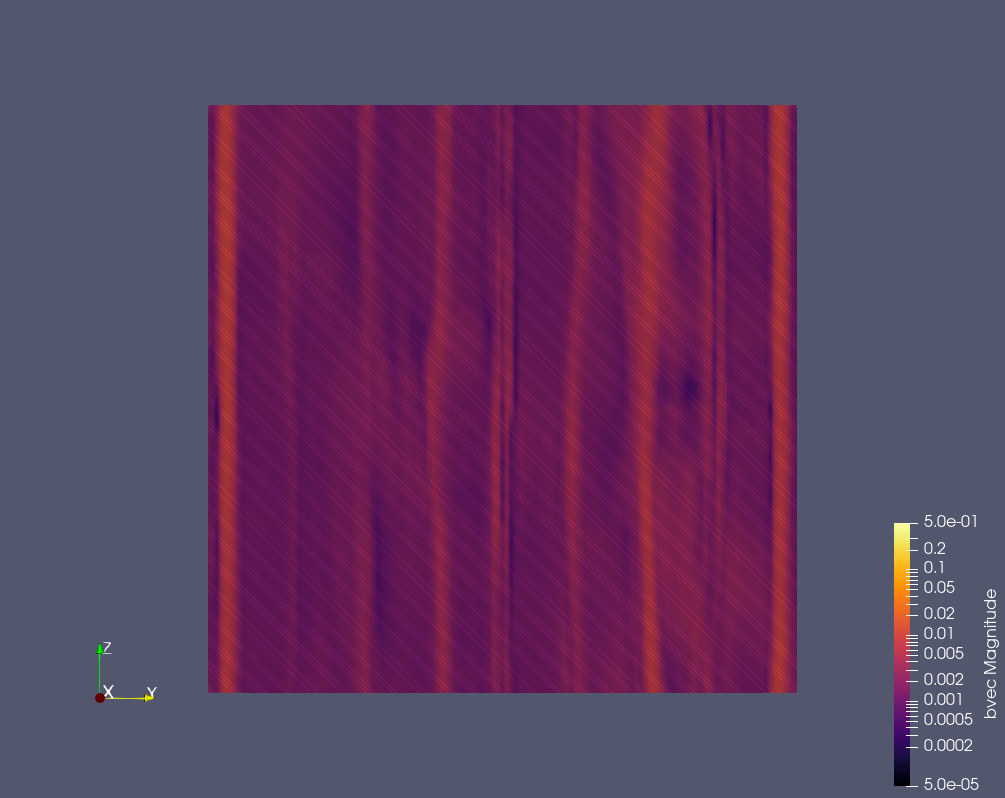} 
\includegraphics[width=0.19\textwidth]{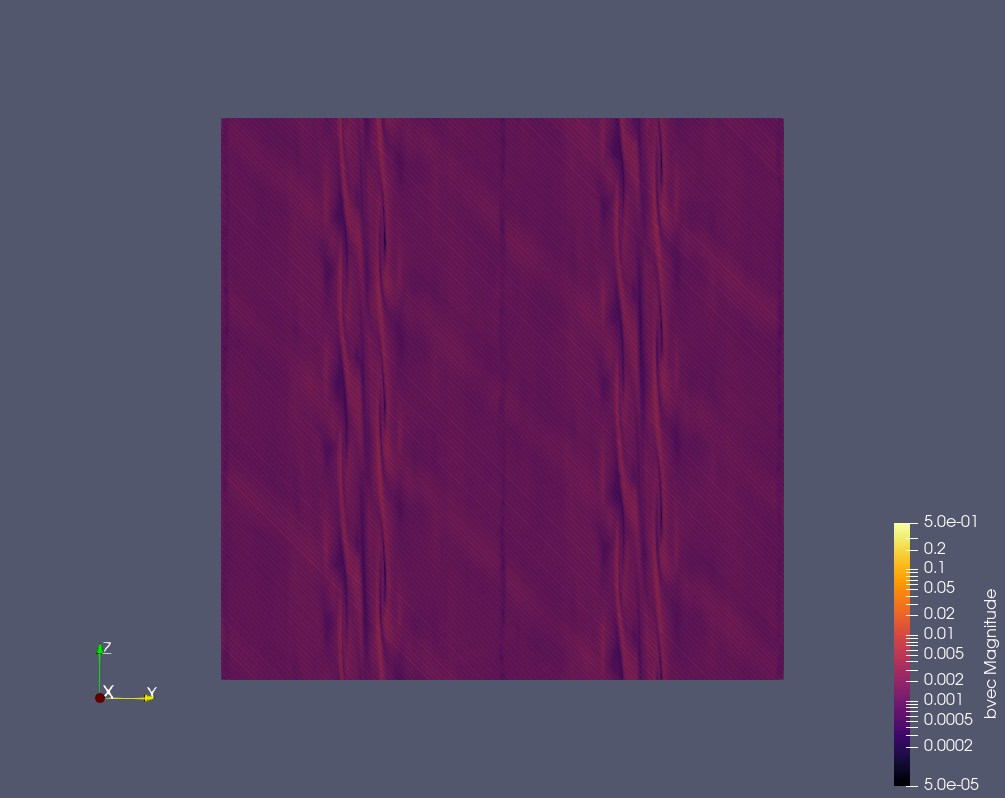}\\
	
	\includegraphics[width=0.19\textwidth]{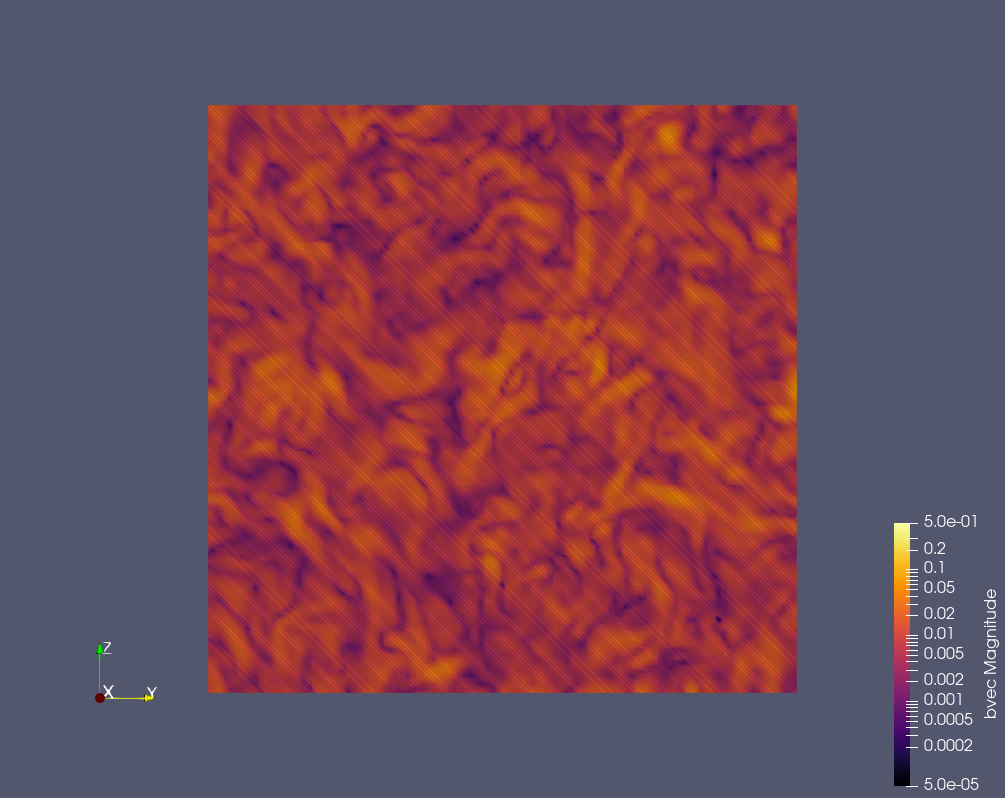} 
\includegraphics[width=0.19\textwidth]{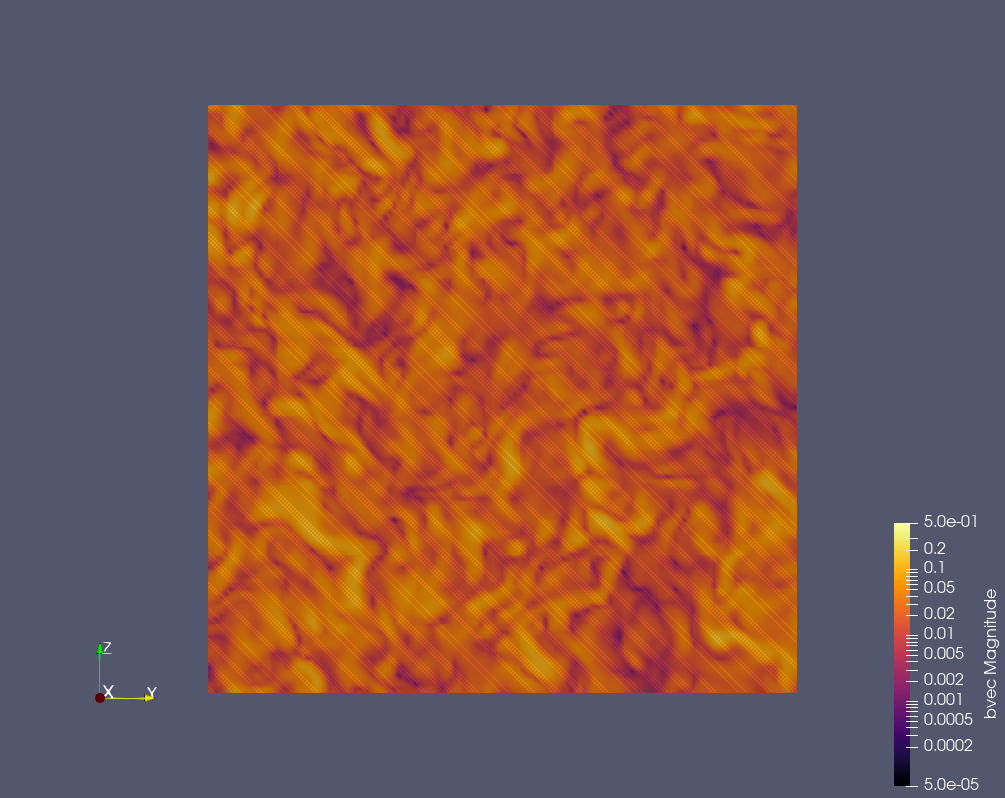} 
\includegraphics[width=0.19\textwidth]{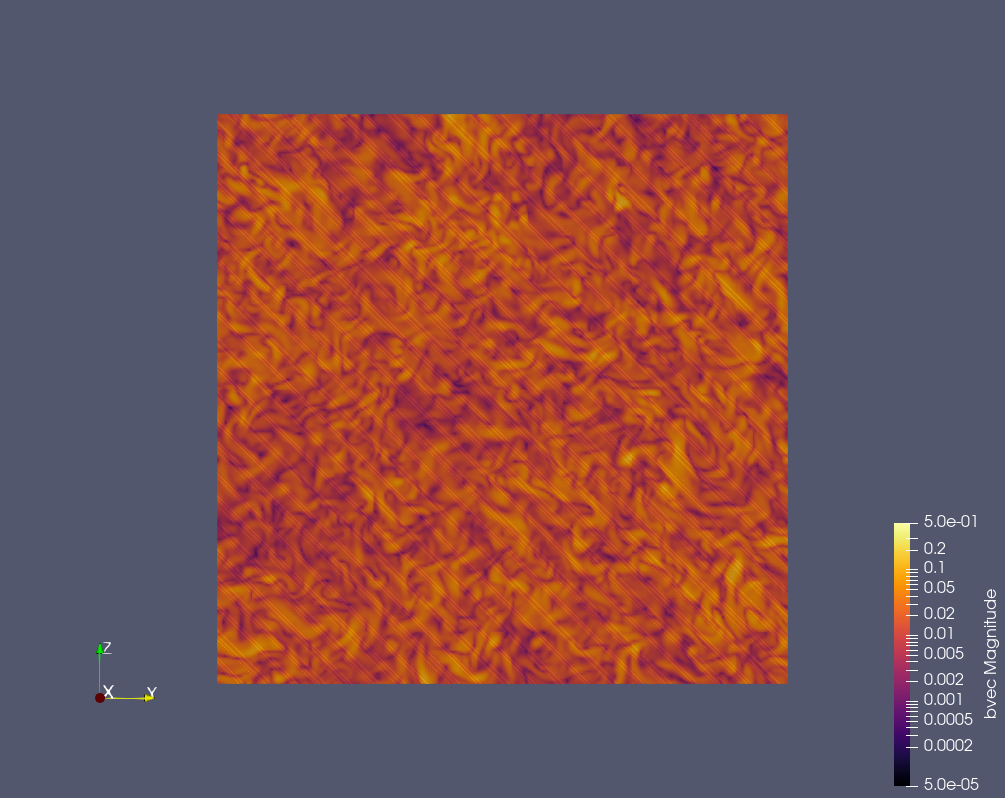} 
\includegraphics[width=0.19\textwidth]{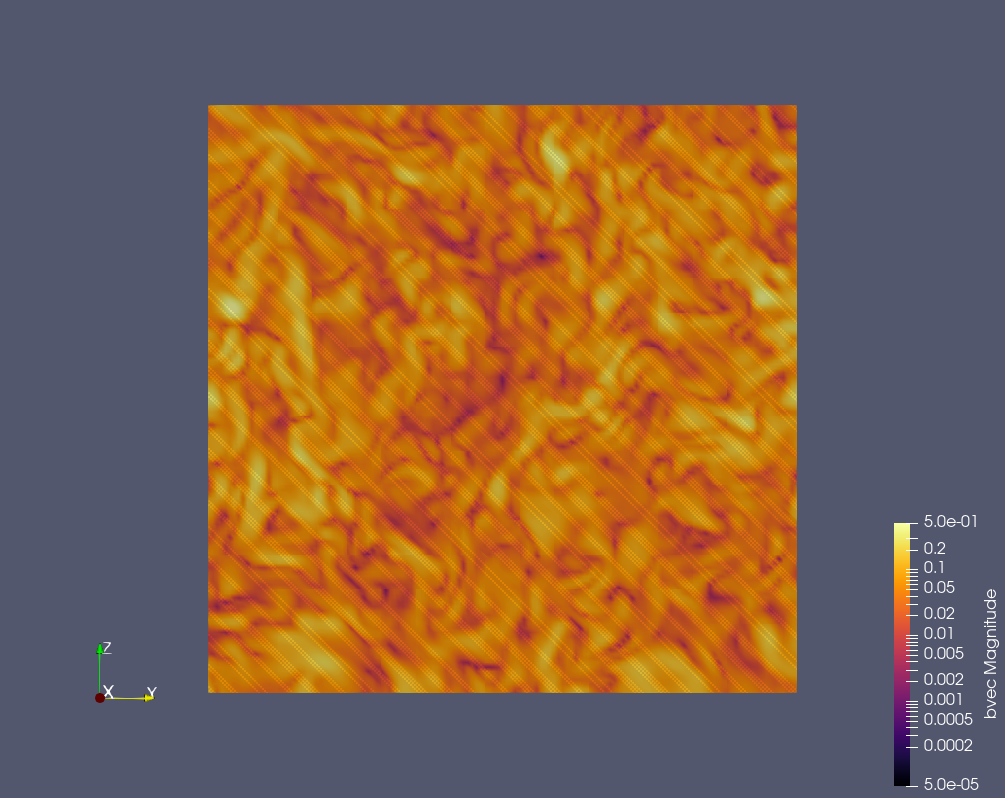} 
\includegraphics[width=0.19\textwidth]{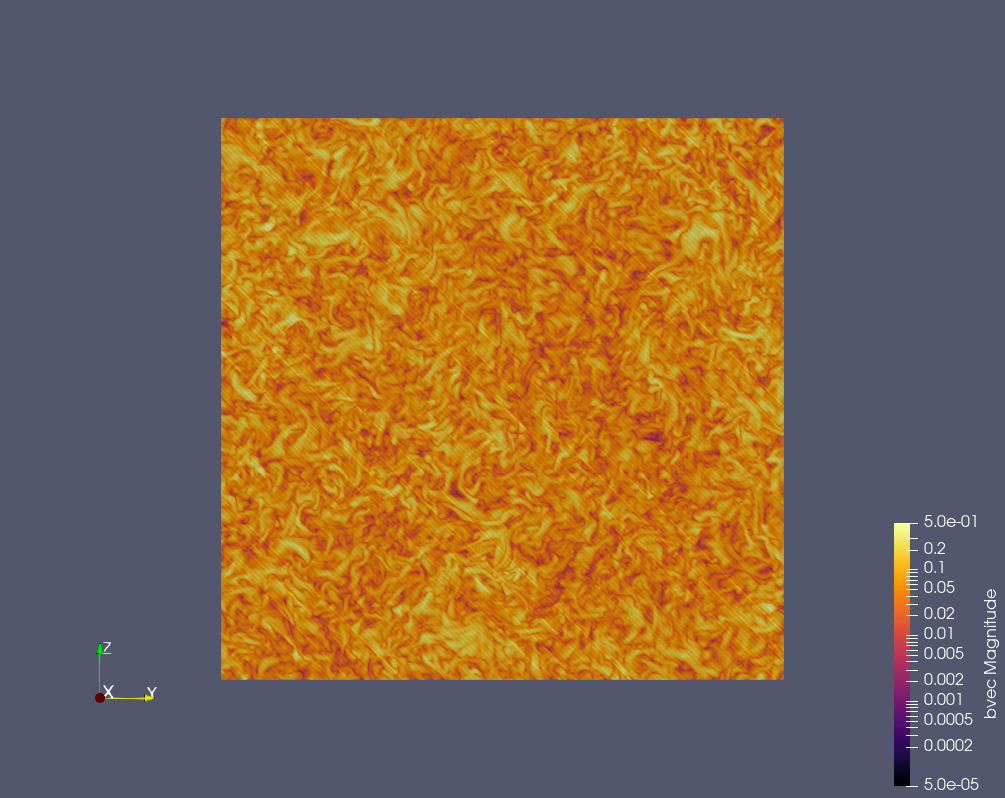}\\
	\includegraphics[width=0.19\textwidth]{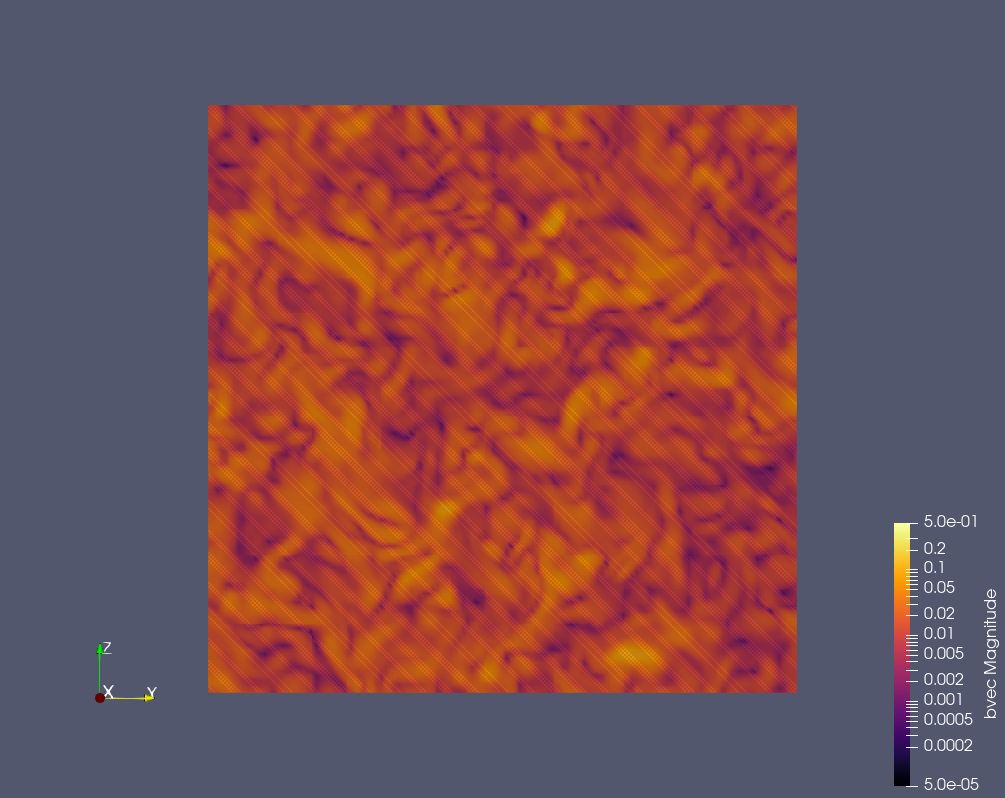} 
\includegraphics[width=0.19\textwidth]{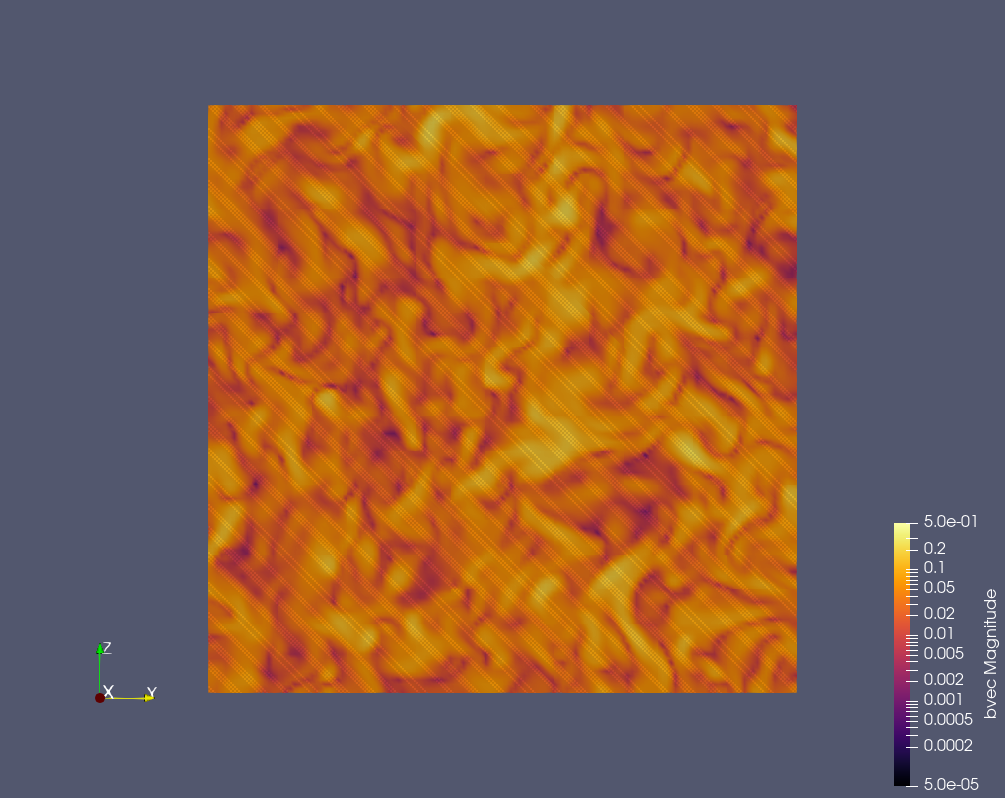} 
\includegraphics[width=0.19\textwidth]{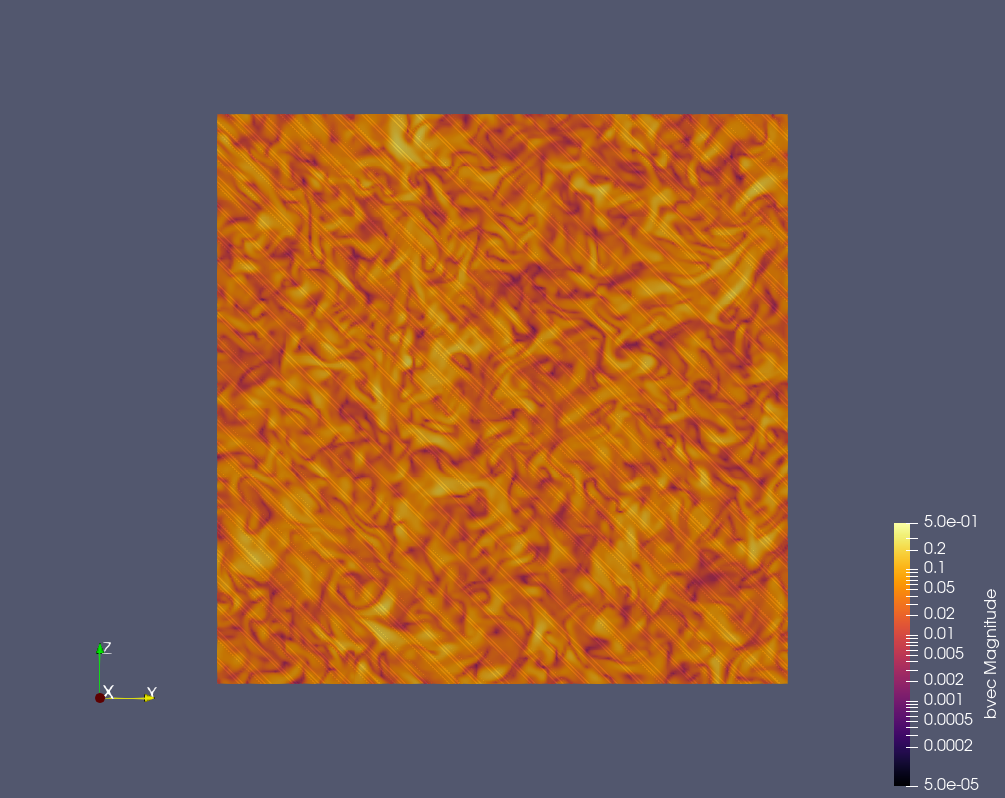} 
\includegraphics[width=0.19\textwidth]{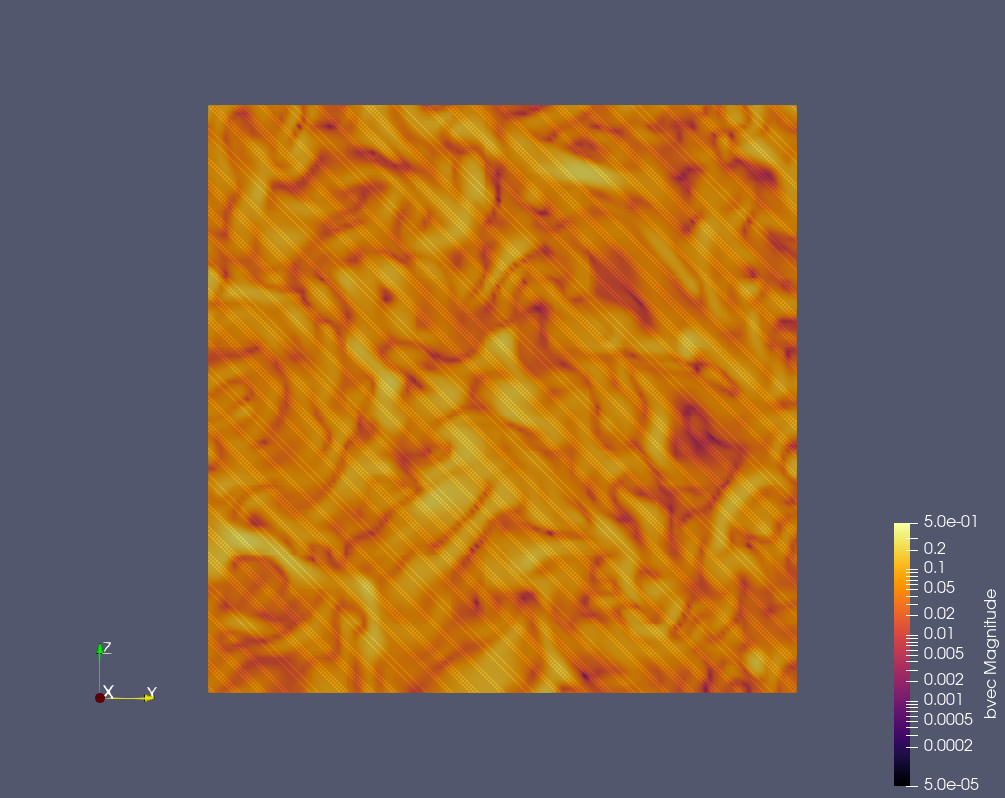} 
\includegraphics[width=0.19\textwidth]{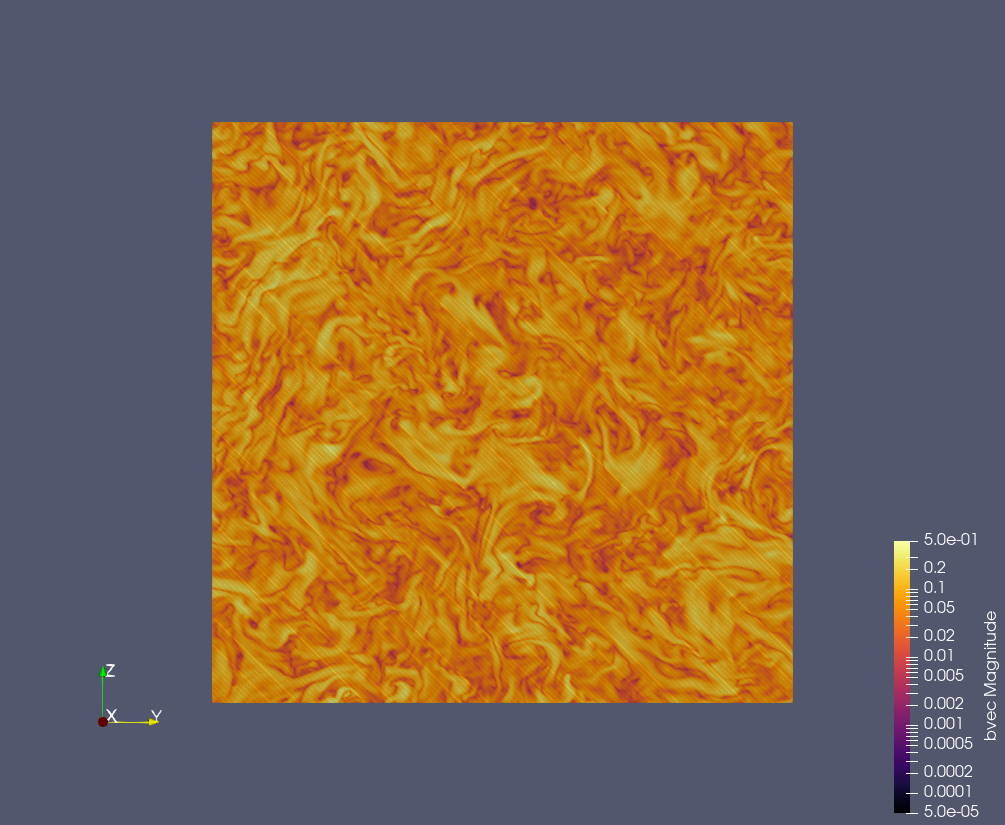}  	
	\caption{ 2D slices of 3D evolution of $|B|$ for special relativistic KHI evolution. The three rows demonstrate different timeslices $t = (0.25,10,20)$. The five columns show the models LR, \texttt{f2S1}, SR, \texttt{f4S1}, HR respectively. }
	\label{fig:sr3d_b}
\end{figure*}

In Figs. \ref{fig:sr3d_rho} - \ref{fig:sr3d_b} we show 2D slices of the full 3D evolution for models LR, SR and HR and two subgrid models \texttt{f2S1}, \texttt{f4S1} at 3 characteristic timeslices.  These are oblique spatial slices taken to show the full 3D evolution, which do not align with the $z$ axis. 

In Fig. \ref{fig:sr3d_rho} we see the evolution of the density. By the middle timeslice $t=10$ we see that the initial conditions have been washed out and that a turbulent state has developed. By the end of the simulation we see that structure cascades to smaller length scales, best captured at high resolution.

In Fig. \ref{fig:sr3d_b} we demonstrate the evolution of the magnetic field. We see that, while the SR and HR simulations are clearly capable of capturing the small scale features in the magnetic field that cannot be directly resolved in any of the LR simulations, the magnitude of the magnetic field, as shown by the colour scale, is comparable between the SR and \texttt{f2S1} runs, and between the HR and \texttt{f4S1} runs. 

\begin{figure}[t]
  \centering 
  \includegraphics[width=0.49\textwidth]{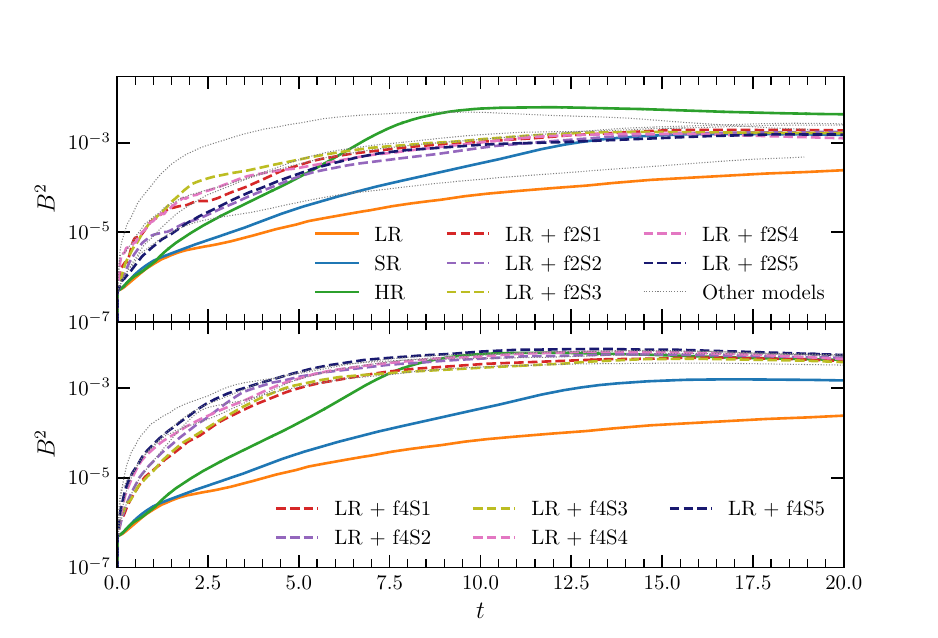} 
  \caption{Amplification of the magnetic field with subgrid model in 3D special relativistic KHI test. Solid lines denote fiducial runs with no subgrid model. Dashed lines show the subgrid models described in Tab. \ref{tab:sr3dhyp}. Grey dotted lines are other models in hyperparameter space Upper panel: \texttt{f2} models trained on SR data. Lower panel: \texttt{f4} models trained on HR data.}
  \label{fig:sr3dbamp}
\end{figure}

In Fig. \ref{fig:sr3dbamp} we directly demonstrate the amplification of the magnetic field. In the upper panel we demonstrate the performance of the \texttt{f2} models, which aim to reproduce the behaviour of the SR run. We see that the four selected models all demonstrate qualitatively similar behaviour. At early times the models amplify faster than their target run, which is still dominated by the impact of the initial data, but that between $t\sim 2.5 -10$ the simulations evolve with a similar growth rate of the magnetic field. At $t\sim 10$ the simulations saturate at a similar magnetic field strength. We detail the final field strength in Tab. \ref{tab:sr3dhyp}, which we find to be consistent with the SR run to between  3 and 27\%. Comparing the speed of evolution we see that these configurations are between $\sim 3-5$ times faster than the SR configuration. In the lower panel we repeat this analysis for the \ff models. The same overall behaviour is seen, a fast initial growth before saturating at the target HR value. The final error in the magnetic field is also between 3-27\% with a speed-up between a factor of 28-44. As above, we have selected from our trained models those that perform well in the a posteriori testing. However we demonstrate in dotted grey lines a broader set of other models that also perform well. This demonstrates that well performing models are not rare in hyperparameter space. Unlike in the Newtonian 2D case we do not denote these as slow models since in the more computationally taxing special relativistic 3D case, all models are faster than the SR or HR simulations.

In all models we consider, the amplification appears earlier than in the target no-subgrid run. We ascribe this behaviour to the fact that we train our models using timeslices from the end of the simulation, to avoid carrying the imprint of the initial conditions. Therefore the NN is biased towards generating the final turbulent state as quickly as possible. For future applications we will be interested in applying this technique in real physical scenarios, therefore we are not interested in training models that perform well in the early phase where the evolution is dominated by the initial data from which the KHI develops. 

\begin{figure}[t]
   \centering 
   \includegraphics[width=0.49\textwidth]{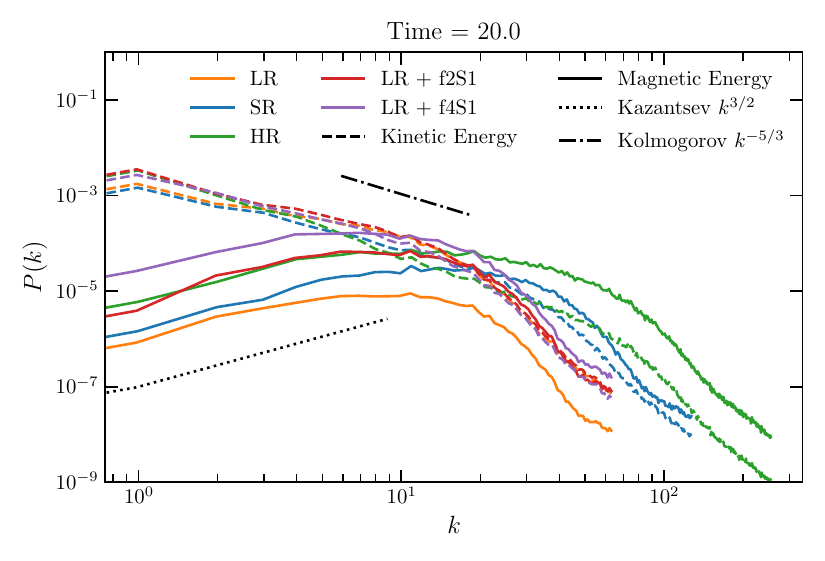} 
   \caption{Spectrum at $t=20$ of magnetic (solid lines) and kinetic (dashed lines) energies. For visualisation purposes the Kolmogorov $k^{-5/3}$ and Kazantsev $k^{3/2}$ scalings are shown with dot dashed lines. Different models are denoted by different colours. }
   \label{fig:sr3dspec}
 \end{figure}

To measure the accuracy of the model we must validate that it not only captures the final amplification of the magnetic field associated to the higher resolution simulation, but also that the structure of the magnetic field is consistent with this simulation. In Fig. \ref{fig:sr3dspec} we demonstrate the power spectrum of the magnetic and kinetic energies. In the presence of the turbulence induced by the KHI, we expect to see that a Kolmogorov cascade emerges in the kinetic energy, demonstrating the well known $k^{-5/3}$ power law scaling \cite{Kolmogorov:1941}. Associated to this cascade of kinetic energy, we expect to see a dynamo process emerging, associated to a $k^{3/2}$ power law scaling following Kazantsev \cite{Kazantsev:1968a}. We see that, by the end of our simulation, all runs demonstrate power law scaling in the kinetic spectra consistent with the Kolmogorov scaling, and in the magnetic field consistent with the Kazantsev scaling. In particular, this verifies that the subgrid models successfully reproduce the dynamo process at work. Clearly, due to the smaller length scales they resolve, we see that the SR and HR simulations capture the magnetic field structure at larger values of $k$ better than the subgrid simulations, and that the larger magnetic field strength in the SR and HR compared to LR simulation is associated mostly to the magnetic field that is captured at smaller scales, and to a lesser extent, the larger magnetic field that develops on longer wavelengths through the dynamo process. In contrast, the subgrid models cannot directly capture the magnetic field on these smaller scales as they are limited by resolution. Therefore, in order to match the magnetic amplification of SR (HR), \texttt{f2(4)S1}  must achieve more amplification at longer wavelengths, as seen in Fig. \ref{fig:sr3dspec}. Then in comparison to LR, model \texttt{f2(4)S1} shows considerable amplification at all wavelengths, with the overall shape of the spectra largely unchanged. We see that the kinetic spectra are largely similar between all models with, again, the subgrid models capturing more kinetic energy at long wavelengths.

 Our a posteriori selection criterion for well performing models was based on matching the final magnetic field amplitude of the target run. Nevertheless, our subgrid models appear to show a better match to the Kolmogorov power law behaviour for the kinetic energy than the LR simulation, even though no information about the kinetic energy was used to select them as well performing models. This suggests that our selection criterion focusing on magnetic field amplification is robust in choosing models that capture all aspects of subgrid physics, not just magnetic field dynamics.

We finally assess the performance of our models by analysing the growth rate of the KHI. We have noted that our subgrid models grow their magnetic fields earlier than the fiducial runs with no subgrid model, due to the fact that we train on late time data. However, it is important that the growth rate of the instability is preserved, as this contains physically important information. In a DNS fully capturing the KHI, we expect the growth of the magnetic field energy to be exponential in time, with a growth rate related to the local behaviour in the vicinity of the shear layer, $B^2 \propto e^{\gamma_{\mathrm{KHI}}t} $
As the instability develops and a fully turbulent state is realised, it is practically difficult to analyse the width of a given shearing layer, though diagnostic proxies for this have been constructed, such as the local shear \cite{Miravet-Tenes:2023see}. In a simulation with finite resolution that cannot directly capture the full cascade of the KHI, it has however been shown that the growth rate should be inversely proportional to the grid spacing $\gamma_{\mathrm{KHI}} \propto \frac{1}{\Delta x}$  \citep{Kiuchi:2015sga}. We model the growth of the magnetic field energy ($E_B$) therefore with the form 
\begin{equation}
E_B = E_0 e^{\gamma_{\mathrm{KHI}}(t - t_0)},
\end{equation}
where $t_0$ is the time at which the KHI begins to grow, after the initial growth associated to the initial conditions, seen uniform across all resolutions in runs without a subgrid model, has finished. We fit this functional form to the data in Fig. \ref{fig:sr3dbamp} for a range of values of $t_0$ to search for the optimal window during which the KHI is active for each model, and extract $\gamma_\mathrm{KHI}$ for each simulation. We then make the assumption that $\gamma_{\mathrm{KHI}}=\gamma / \Delta x$, where $\gamma$ is a universal constant. From our 3 configurations without a subgrid model (LR, SR, HR) we extract an average value of $\gamma = 0.00408$. This value allows us to extract an ``effective'' $\Delta x$, or equivalently an effective number of grid points $N_{\mathrm{mesh,eff}} = 2 \gamma_{\mathrm{KHI}}/\gamma $ for each of our runs employing a subgrid model, by comparing this value of $\gamma$ to the $\gamma_\mathrm{KHI}$ fitted for each subgrid run. This demonstrates the resolution that would be needed to obtain an equivalent growth rate of the KHI, seen instead at low resolution with the subgrid model employed. Our results are detailed in Table \ref{tab:sr3dhyp}, and we see that, of the \ft models, the effective resolution largely matches the target SR resolution of 256 points, with three of our models below 12\% error, with two other models performing notably worse with error at worst 38\%. For the \ff models, we see slightly better performance in recovering the target growth rate with three models showing under $6\%$ error, and the remaining two showing at worst an error of $20\%$.

\subsubsection{Robustness of models with $\tau_D$}\label{sec:sr_individ}

In the models presented above we have included the impact of all subgrid terms, including that which enters the right hand side of the density evolution equation $\tau_D$. In our a posteriori tests, we have empirically found that it is easier to find successful models in hyperparameter space that match the target magnetic field amplification by deactivating this term, and that, with this term active, there is a larger set of hyperparameters that lead to models that over-amplify the magnetic field. This over amplification can drive the simulation to unphysical values and eventually lead to a sufficiently strong magnetisation that the Ideal MHD equations become ill-conditioned, leading to the crashing of our numerical code. Nevertheless we find a sufficiently large region of hyperparameter space containing models that give successful evolutions with this term active, to conclude that robust and accurate models and evolutions can be constructed with all subgrid terms active.

The impact of this term in the density evolution equation is relatively rarely studied in the literature. In the case of Newtonian MHD, due to the particular choice of the Favre filter operation on the fluid velocity this subgrid term can be removed by definition. Therefore, in the extensive literature on non-relativistic (M)HD LES simulations, this term can be neglected. In the relativistic case, due to the non-linear dependence of the momentum density on the fluid 3-velocity, through the presence of the Lorentz factor $W = \frac{1}{\sqrt{1 - v^2}}$, this choice of filter does not allow a simple elimination of this subgrid term. Of the other authors who have focused on LES simulation in fully relativistic MHD, only in the gradient approach of \cite{Carrasco:2019uzl,Vigano:2020ouc}, have the full spectrum of terms, including this term in the density evolution, been considered. 

In their study of a posteriori performance (in GRMHD), \citet{Vigano:2020ouc} found that, with the addition of an extra prefactor of 8 which multiplies the overall subgrid term, tuned to data from a KHI simulation; an a posteriori test with all subgrid terms active reproduces the magnetic field behaviour seen  at double the resolution. It is shown that within this model the dominant effect on the magnetic field comes from the subgrid term in the induction equation only, with all other subgrid terms switched off. As discussed above in the Newtonian case (Sec. \ref{sec:newt_termbyterm}), we have found that in our model, all subgrid terms are important in recovering the high resolution behaviour. In full production applications, in the context of a BNS merger \cite{Aguilera-Miret:2020dhz}, the same group has found that magnetic field amplification is enhanced by increasing the prefactor on the subgrid term in the induction equation but is suppressed by introducing a prefactor for the terms appearing in the momentum and density equations, and that these terms can strongly influence the collapse time of a post merger remnant to a black hole. The conclusion of their study is that to capture the behaviour of the higher resolution simulations without a subgrid model, all subgrid terms other than that in the induction equation should be deactivated, with the term in the induction equation premultiplied by a factor 8, the approach taken in subsequent works \cite{Palenzuela:2021gdo,Aguilera-Miret:2021fre}. Here we have shown that, for the test problems considered, the high resolution behaviour can be matched with all subgrid terms active, including that in the density equation, without the addition of a prefactor, though we note that our model is, similar to this prefactor, entirely fitted to data. It is also the case that the problems we consider here are far more idealised than a full BNS simulation. In a full application of this approach to a BNS merger, it will be necessary to re-examine the relative importance of these terms, and the robustness of our scheme to this problem. We may hope however, that the inclusion of terms in the momentum equation especially, allow us to accurately capture the impact of the small scale magnetic field on momentum transport in the system.

\section{Conclusion}
\label{sec:con}

In this paper we have presented the first implementation of an online subgrid model, tested a posteriori, for modelling turbulent flow in relativistic MHD using NNs. We have demonstrated the application of this model in two test problems of the KHI, firstly a 2D problem in Newtonian MHD, and secondly a fully 3D problem in SR.

In both cases we initially demonstrated that we can train relatively small NNs, on high resolution data, filtered down to lower resolution, to successfully evaluate subgrid tensors, using the fluid state and its derivatives as inputs. We have shown that these models can reproduce a training dataset with good accuracy, passing the initial a priori test that such subgrid models have previously been subjected to in \cite{Rosofsky:2020a}.

We have then gone beyond these tests to, for the first time, show success in a posteriori testing. To demonstrate this we have performed low resolution simulations, evaluating the NN online in the simulation to provide the subgrid terms. We have shown that models can be constructed that successfully reproduce the high resolution total amplification of the magnetic field, Kelvin-Helmholtz growth rate, and spectral distribution of magnetic and kinetic energies in these low resolution simulations, with a speed up of a factor up to 44. 

Further, we have investigated the role of individual subgrid terms in the evolution of such problems. In the Newtonian case we have seen that subgrid models that successfully give the expected magnetic field amplification in a posteriori tests do so due to the impact of all relevant subgrid terms, incorporating the impact on the momentum, energy and magnetic field evolutions. This suggests that subgrid models that only include the impact on e.g. the magnetic field evolution, may neglect important contributions to the magnetic field evolution. We have also seen in the SR case that robust models in a posteriori testing can be found even when the subgrid contribution to the density evolution, absent in Newtonian physics, is included, providing a model that incorporates all subgrid effects at once.

This work demonstrates the applicability of such techniques to more complicated scenarios, specifically full general relativity, in the context of isolated neutron star evolution, binary neutron star mergers and accretion flows around a black hole. Future work will investigate applying such techniques to these configurations and modelling the magnetic field evolution at relatively low resolution with the use of such subgrid models. Our work in this paper has focused solely on the KHI, however other authors constructing different subgrid models have focused on constructing bespoke models for separate magnetic field instabilities that may trigger turbulent flow, specifically both the KHI \cite{Miravet-Tenes:2023see} and the MRI and associated parasitic instabilities  \cite{Miravet-Tenes:2022ebf} . Future work is required on our ML based model to understand whether bespoke NNs are required to capture each of these separate instabilities or whether a single network can be trained to reproduce all of the relevant subgrid physics at once. 

The filtering operations discussed in this work break the covariance of general relativity. It has been suggested that this breaking of covariance should lead to gauge artefacts which impact the thermodynamics of the evolved system \citep{Celora:2021fry,Celora:2024rif,Celora:2024iqr}. When considering the general relativistic extension to this problem, it may prove important to implement the above subgrid approach in a gauge invariant manner to obtain robust physical results.

\begin{acknowledgments}
  The authors thank Charles Board, Thomas Celora and Prasoon Pandey for helpful conversations. 
  SB acknowledges support by the EU Horizon under ERC Consolidator
  Grant, no. InspiReM-101043372.
  SB acknowledges support by the DFG project ``Magnetfelddynamik in
  Neutronensternen Sternen'' MERLIN (MERLIN; BE 6301/6-1
  Projektnummer: 524726453).

  Simulations were performed on the national HPE Apollo Hawk (Hunter)
  at the High Performance Computing Center Stuttgart (HLRS).
  The authors acknowledge HLRS for funding this project by providing access
  to the supercomputer HPE Apollo Hawk (Hunter) under the grant numbers INTRHYGUE/44215
  and MAGNETIST/44288.
  Simulations were also performed on SuperMUC\_NG at the
  Leibniz-Rechenzentrum (LRZ) Munich.
  The authors acknowledge the Gauss Centre for Supercomputing
  e.V. (\url{www.gauss-centre.eu}) for funding this project by providing
  computing time on the GCS Supercomputer SuperMUC-NG at LRZ
  (allocations {\tt pn67xo}, {\tt pn76li}).
  
    Simulations and post-processing were also performed on the ARA and DRACO clusters
  at Friedrich Schiller University Jena and on the {\tt Tullio} INFN cluster at INFN Turin.
  The ARA cluster is funded in part by DFG grants INST
  275/334-1 FUGG and INST 275/363-1 FUGG, and ERC Starting Grant, grant
  agreement no. BinGraSp-714626.
\end{acknowledgments}

\appendix

\section{Hyperparameter optimisation and training data set construction}
\label{app:hyp}

In order to find performant, accurate models, we train models with a wide range of hyperparameters.
We assess the performance of these models both in a priori and a posteriori tests, and have presented the 
best performing models in the analysis above. In particular we iterate over the number of hidden layers $N_L$, the number of neurons per hidden layer $N_n$, and the choice of non linear activation function. As well as the hyperparameters of the model architecture,
we have also iterated over various methods of dataset construction. For the Newtonian 2D problem this entails exploring strategies for augmenting the tails of the dataset which is comparatively small due to the dimensionality of the problem, while in the 3D special relativistic case, this has involved subsampling the spatial domain included in the training set.

\begin{figure}[t]
  \centering 
  \includegraphics[width=0.49\textwidth]{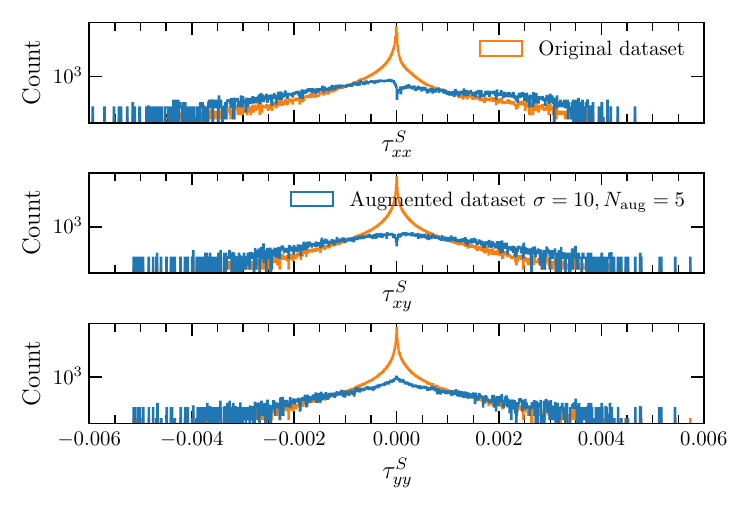} 
  \caption{{ Histograms of the training dataset for the components of the momentum subgrid tensor $\tau^S_{ij}$ before (orange) and after (blue) the data augmentation procedure. This dataset was used to train model N3.}}
  \label{fig:dataaug}
\end{figure}

Our data augmentation strategy for the 2D case is driven by the fact that the region of most importance for our subgrid model is the tail of the distribution of subgrid tensor values within our training set, that is, we are most interested in accurately modelling regions where the subgrid tensor takes large values, which are rare within our training set. We therefore pursue a strategy that boosts the tails of the distribution and suppresses the mean. To do this we take our unaugmented training data set and calculate a statistic for each value of a subgrid tensor component $\tau$,
\begin{eqnarray}
	\tau_\mathrm{remove} &=& 1 - (1-10^{-6})\exp\left(\frac{-\tau^2}{2(\sigma \sigma_\tau)^2}\right)\\
	\tau_\mathrm{add} & = &  1 - (1-10^{-6})\exp\left(\frac{-\tau^2}{2(2\sigma \sigma_\tau)^2}\right)  ,
\end{eqnarray}
where $\sigma_\tau$ is the standard deviation of the distribution of $\tau$ in the unaugmented dataset. The mean of all distributions considered is $\approx 0$. 
We then generate a random variable drawn from a uniform distribution $\tau_\mathrm{cut} \sim \mathrm{Unif}[0,1]$ for each element of our dataset and compare it to these statistics. If $\tau_\mathrm{cut} > \tau_\mathrm{remove} $ we remove the point from our dataset, and if  $\tau_\mathrm{cut} < \tau_\mathrm{add} $ we add $N_\mathrm{aug}$ copies of the point to the dataset; thus flattening the distribution.  $\sigma,N_\mathrm{aug}$ are hyperparameters we search over. In Fig. \ref{fig:dataaug} we show the impact of this data augmentation on the training data.

 We can see that the unaugmented dataset is heavily weighted towards the mean, where the value of the subgrid tensor is small, representing non-turbulent regions within the simulation. Since we require small models for fast evaluation to retain performance, this unaugmented dataset will lead to a model well trained for non turbulent regimes, but with poor performance in the turbulent state. These configurations can correspond to those discussed above, which perform well in a priori tests, but poorly in a posteriori tests. In comparison, we find the flattened, augmented dataset, to give models better trained in all physical regimes of our problem. This augmentation strategy is stochastic, however we perform multiple realisations of the augmented dataset and verify that the resulting dataset does not significantly vary.

\begin{figure}[t]
  \centering 
  \includegraphics[width=0.49\textwidth]{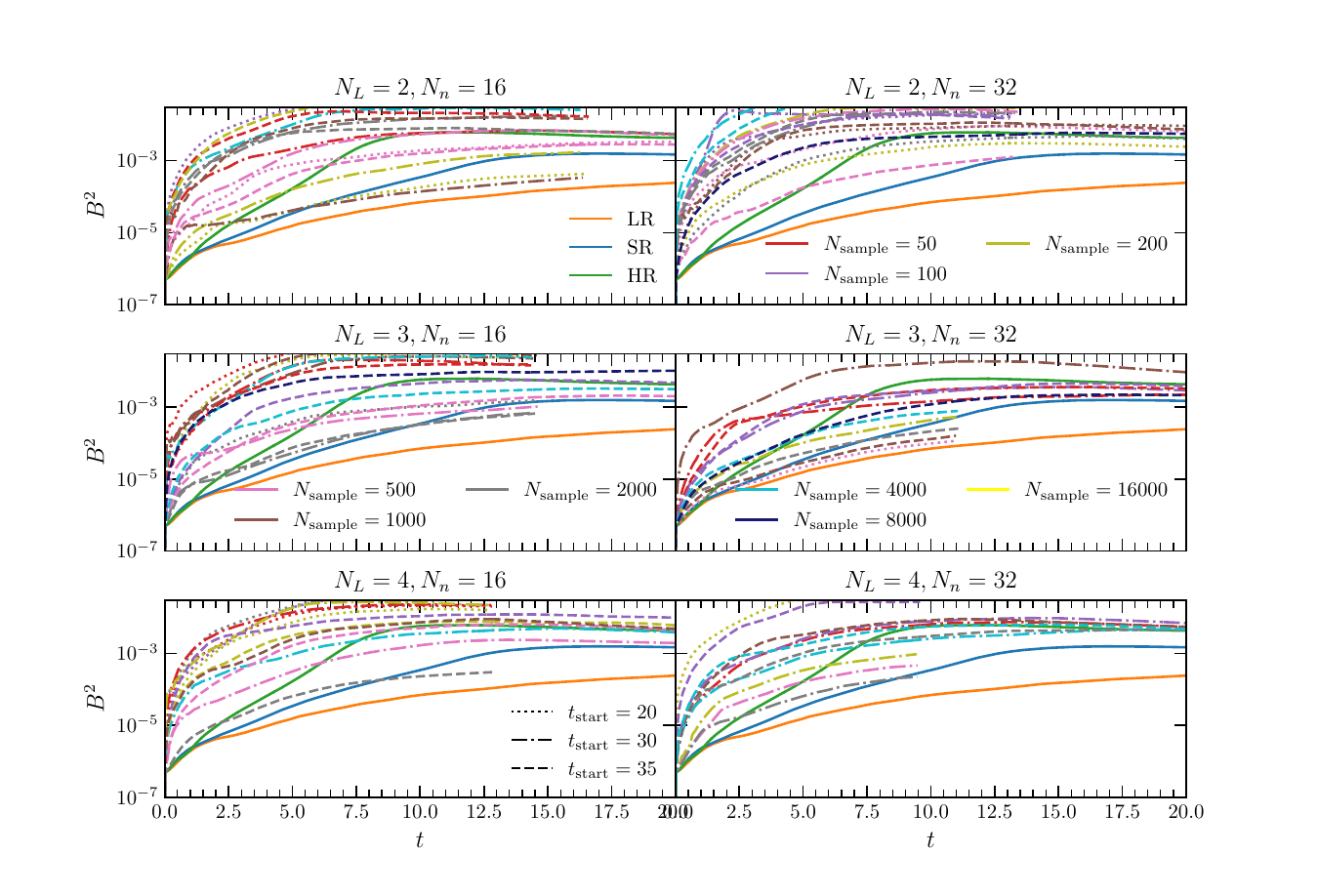} 
  \caption{{ Performance of a range of \ff models spanning a subset of hyperparameter space. Solid curves show the reference runs without subgrid models. Different sub-panels show different choices of the number of hidden layers $N_L$ and number of neurons $N_n$. Line colour shows the choice of sampling parameter $N_{\mathrm{sample}}$, while line style shows the choice of $t_\mathrm{start}$. }}
  \label{fig:hypersearch}
\end{figure}

The range of hyperparameters we iterate over is given in Table \ref{tab:hyper}.
We have selected network sizes to ensure that the evaluation time of the network online in the a posteriori test
is not so slow that a low resolution run with the network is slower than a high resolution run without the network. In total we have explored 390 configurations for the Newtonian problem and 347 for the SR problem, run with full a posteriori tests. From this wide span of hyperparameter space we have shown the best five performing models in each family in a posteriori tests above and a small subset of the other successful models. We note that many models in the Newtonian case without data augmentation show insufficient amplification of the magnetic field, as do models trained on too early a set of timeslices which do not capture the late time turbulent state sufficiently well. In Fig. \ref{fig:hypersearch} we show a wider subset of hyperparameters explored for the special relativistic problem, focusing on the \ff models.  This is a subset of the full hyperparameter set which would be impractical to show in its entirety. Note that some simulations were terminated early if the evolution was clearly not trending towards the targeted growth of the magnetic field. We also note that some simulations not pictured show such a strong growth in the magnetic field that unphysical values are reached and the code generates NaNs. We see here however that well performing models are not a rare outcome of the training procedure, but also that a spread of outcomes can be expected in terms of growth rates and magnetic field amplifications, and that such a posteriori test are necessary to select the best performing models.

\begin{table}[t]
  \centering    
  \caption{Hyperparameter ranges. (N) denotes a parameter only relevant to the Newtonian case, (S) for the SR case.}
  \begin{tabular}{c|c}        
    \hline
    Hyperparameter & Values\\
    \hline
    $N_L$ & 2, 3, 4, 5, 6\\
    $N_n$ & 16, 32, 64 \\
    Act. Func. & $\mathrm{ReLU}(x), \tanh(x), \mathrm{Leaky ReLU}(x)$\\
    Learning Rate & $10^{-3}, 10^{-4}$\\
    Batch size & 1000\\
    $t_{\mathrm{start}}$ (N) & 50, 75 \\
    $t_{\mathrm{start}}$ (S) & 20, 30, 35 \\
    $\sigma$ (N)& 0, 5, 10, 15, 20, 25\\
    $N_{\mathrm{aug}}$ (N)& 0, 5, 10, 15, 20, 25\\
    $N_{\mathrm{sample}}$ (S) & 50, 100, 200, 500, 1000, 2000, \\& 4000, 8000, 16000\\
    \hline
  \end{tabular}
  \label{tab:hyper}
\end{table}

\end{document}